# Metrics for Quantifying Shareability in Transportation Networks: The Maximum Network Flow Overlap Problem

**Navjyoth Sarma J S** [a,b]

Email: nsarma.js@uci.edu

ORCiD: 0000-0002-1304-0162

**Michael Hyland** [a,b] (Corresponding Author)

Email: hylandm@uci.edu

ORCiD: 0000-0001-8394-8064

4000 Anteater Instruction & Research Bldg., Irvine, CA 92697-3600

[a] University of California-Irvine, Institute of Transportation Studies

[b] University of California-Irvine, Department of Civil and Environmental Engineering

July 13, 2022

# ABSTRACT

Cities around the world vary in terms of their urban forms, transportation networks, and travel demand patterns; these variations affect opportunities for travelers to share trips, and the viability of shared mobility services. This study proposes metrics to quantify the maximum shareability of person-trips in a city, or region of a city, as a function of two inputs—the transportation network structure and origin-destination (OD) travel demand. The study first conceptualizes a fundamental shareability unit, 'flow overlap'. Flow overlap denotes, for a person-trip traversing a given path, the weighted (by link distance) average number of other person-trips sharing the links along the original person-trip's path. The study extends this concept to the network level and formulates the Maximum Network Flow Overlap Problem (MNFLOP) to assign all OD person-trips to network paths that maximize flow overlap in the whole network. The study also proposes an MNFLOP variant with a second objective function term, detour distance, to capture the trade-off between minimizing travel distance and maximizing shareability. The study utilizes the MNFLOP output to calculate metrics of shareability at various levels of aggregation: person-trip level, OD level, origin or destination level, network level, and link level. The study applies the MNFLOP and associated shareability metrics to different OD demand scenarios in the Sioux Falls network. The computational results verify that (i) MNFLOP assigns person-trips to paths such that flow overlaps significantly increase relative to shortest path assignment, (ii) MNFLOP and its associated shareability metrics can meaningfully differentiate between different OD trip matrices in terms of maximum shareability, and (iii) an MNFLOP-based metric can quantify demand dispersion—a metric of the directionality of demand—in addition to the magnitude of demand, for trips originating or terminating from a single node/location in the network. The paper also includes an extensive discussion of potential future uses of the MNFLOP and its associated shareability metrics.

# 1 Introduction

## 1.1 Motivation

Cities are engines of economic growth and opportunity, as they connect individuals of all income levels with opportunities, most notably employment opportunities but also healthcare, social/recreational, and religious opportunities (Glaeser, 2011). As a result of the opportunities they provide, today an estimated 56% of the global population live in cities, with an additional 2.5 billion people estimated to move to cities by 2050 (World Bank, 2019). The high and increasing density of humans in cities has wide-ranging implications for civil infrastructure systems, including urban transportation systems.

In developed countries around the world, most notably in the United States, the personal vehicle is the predominant mode of travel even within large cities. Unfortunately, personal vehicles consume considerable space when they are either parked or driven in dense urban areas. Hence, as cities grow, there is no means to prevent heavy congestion and gridlock if travelers rely on their personal vehicles to travel. Already, the average auto commuter in the US loses nearly 54 hours in congestion per year, wasting nearly $1,100 annually in time and fuel (Schrank et al., 2019).

To provide the large and growing number of urban residents with the mobility that underlies access to opportunities, cities will need to offer, and travelers will need to use, significantly more space-efficient modes of travel than the personal vehicle. Figure 1-1 displays a portfolio of different shared mobility options that cities can utilize to provide people with mobility and accessibility. The options vary along two overlapping dimensions, namely, (i) flexibility in route and schedule and (ii) capacity or space efficiency. On the left are the highest capacity and most space efficient shared mobility modes, including rail transit, and fixed-route, fixed-schedule bus transit. On the far right, are the lowest capacity and least-space-efficient (but most flexible in terms of route and schedule) shared mobility modes. Figure 1-1 also shows that there are a range of shared mobility modes between these two ends of the spectrum.

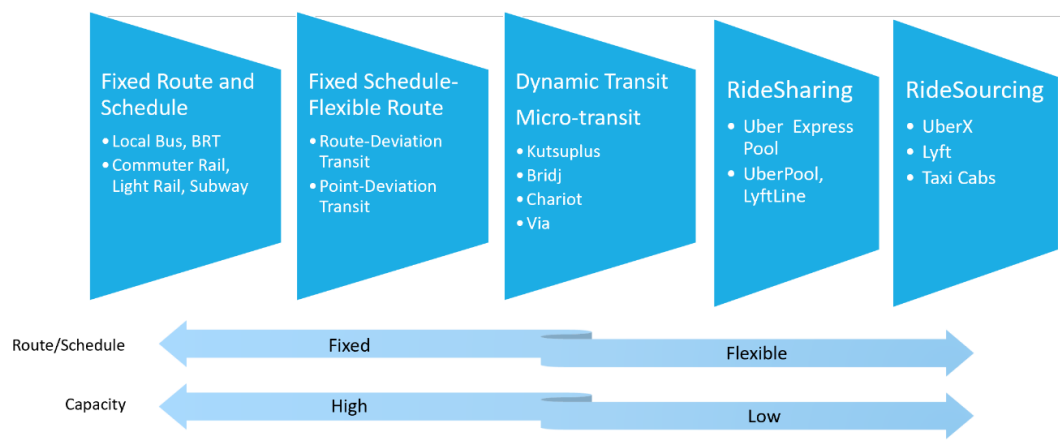

**Figure 1-1 Spectrum of Shared Mobility Modes**

A remaining challenge for cities is determining where (i.e., what subregions within a metropolitan area) and when (i.e., at what times of day) each of these shared mobility options are optimal or even viable. Addressing this challenge is critical to taking advantage of the benefits of shared mobility options and addressing the larger challenge of providing space-efficient mobility in dense urban areas.

Several studies in the literature attempt to answer the 'where' question for flexible transit, or compare fixed and flexible transit options, based on a single parameter, density of demand within a service area (Li and Quadrifoglio, 2010; Nourbakhsh and Ouyang, 2012; Quadrifoglio and Li, 2009). Unfortunately, this single input parameter only captures the magnitude of demand and does not capture the directionality of demand. Moreover, these studies do not consider the underlying transportation network when assessing shared mobility options. Nor do they capture the significant heterogeneity of demand (direction and magnitude) within an urban region. Hence, for researchers and cities interested in determining where and when various shared mobility options are viable, there is a need for metrics that characterize, quantitatively, the magnitude and direction of travel demand while explicitly considering the underlying transportation network structure. These metrics should effectively characterize the key attributes of urban subregions in terms of the subregions' ability to support shared mobility options.

## 1.2   Research Goal, Objectives, and Contributions

The overarching goal of this study is to develop metrics that characterize and quantify sharing mobility potential (i.e., shareability) in urban areas. This study introduces shareability metrics that quantify fundamental urban system and subsystem properties related to both travel demand patterns and street network structure in a city. This study aims to develop shareability metrics that support strategic planning and design decisions, as opposed to tactical or operational planning decisions.

Associated with this overarching goal are five specific objectives. First, the study aims to define and mathematically formulate a fundamental unit of shareability for a person-trip in a transportation network, which we call 'flow overlap'. Second, the study aims to extend this fundamental unit to the network level in order to maximize total network-wide flow overlap. To address this objective, we formulate the Maximum Network Flow Overlap Problem (MNFLOP) that assigns person-trips to paths in a network. Third, the study aims to allow for trade-offs between flow overlap and detour distances when assigning person-trips to network paths via MNFLOP. Fourth, using the MNFLOP output, this study aims to define and formulate a wide range of shareability metrics, at different levels of aggregation in a transportation network, such as OD pair level, location (i.e., node) level, network level and link level. Finally, the study aims to verify (i) the ability of MNFLOP to increase flow overlaps compared to shortest path assignment; (ii) the usefulness of MNFLOP and associated shareability metrics in differentiating between different demand patterns in terms of their sharing potential; and (iii) the usefulness of MNFLOP and a location-level shareability metric to capture the magnitude and directionality of demand emanating from a single node/location while considering the underlying transportation network.

This study makes several contributions to the existing literature. First, we believe this is the first study to define and characterize sharing potential (i.e., shareability) in a mode- and vehicle-independent manner by jointly considering OD person-trip demand, the underlying network structure, and the assignment of person-trip flows through the network. Prior research by Santi et al. (2014) measures shareability, but in the context of a specific mobility service/mode that requires assigning persons to capacitated vehicles. Additionally, while Kucharski and Cats (2020) measure shareability by analyzing demand patterns independent of vehicles, their shareability metrics do not model travel demand as flows through a network. Moreover, their shareability metrics—which include reduction in VMT, number of shareable rides, and total passenger utility—are fundamentally different than the person-trip-based shareability metrics proposed in the current paper. Second, the current study defines and formulates the MNFLOP—a novel mathematical program able to quantify the maximum flow overlap (i.e., this study's measure of shareability) of a network and OD demand, jointly. This formulation, as far as we are aware, represents a novel network flow problem. Moreover, the formulation of the MNFLOP with two objectives—minimizing detour and maximizing overlap—allows researchers and cities to evaluate trade-offs between maximizing shareability (i.e., flow overlaps) and minimizing detours. Third, this study introduces a node- or location-level metric for demand that quantifies both the magnitude and directionality of demand in a transportation network, considering the underlying transportation network, termed *demand dispersion*. Much of the existing literature that aims to characterize demand for shared mobility systems, only considers the magnitude of demand emanating from a single location or network subregion, e.g., Li & Quadrifoglio (2010), Nourbakhsh & Ouyang (2012), and Quadrifoglio & Li (2009). The current study characterizes the degree of demand *dispersion* associated with trips stemming from an origin location, i.e., the extent of commonality or the lack thereof, between all person-trips originating from the same location.

Finally, another valuable feature of the novel MNFLOP model is that it applies to any network with a set of flows, not just transportation networks. Traditionally, constraints and objective functions in network flow models try to avoid, either explicitly or implicitly, condensing flows onto few links, because most network flow applications involve link or node capacity constraints and/or flow-based link or node congestion penalty functions. However, as indicated in this study, there are potential network flow applications where condensing flows onto fewer and fewer links is an appropriate objective.

## 1.3   Paper Structure

The rest of the paper is organized as follows. Section 2 summarizes the relevant literature and explains the gaps this paper attempts to fill. Section 3 introduces a conceptual framework that defines and connects flow overlaps in a network to the notion of shareability. Section 4 employs the definition of flow overlap and the conceptual framework to describe and formulate the Maximum Network Flow Overlap Problem MNFLOP. Section 5 uses the output of the MNFLOP to formulate various shareability metrics at different levels of aggregation in a network. Section 6 provides a summary of the input data for the numerical experiments and scenario analysis. Section 7 presents and discusses shareability metrics for the Sioux Falls network under a variety of scenarios. Section 8 concludes the paper by summarizing the study and discussing limitations and future research.

## 2   Literature Review

### 2.1   Characterizing Urban Form, Transportation Networks, and Mobility Patterns

Several studies in the literature propose methods to characterize urban form/structure and transportation networks. Srinivasan (2002) developed quantifiable neighborhood level characteristics using Geographic Information Systems (GIS) that account for land use, network, and accessibility for the Boston metro area in order to model mode choice for work and non-work tours. The specific metrics include population density, distance to transportation facilities like transit stations, nearest freeway ramps, street density, residential entropy, commercial-residential homogeneity, transport entropy, cul-de-sac density, presence of sidewalks, and width of the sidewalk. The model results indicate that land-use and spatial characteristics of the network and neighborhood significantly impact mode choice. Although Srinivasan (2002) does not characterize shareability for a neighborhood using overlaps in network paths, their study establishes that land-use and the transportation network do affect the usage of shared modes.

Tsai (2005) presents several metropolitan level metrics to characterize different urban forms and distinguish compactness from sprawl. The study uses four distinguishable dimensions of a metropolitan area—namely size, density, degree of equal distribution, and degree of clustering—to define a metropolitan area. The metrics include population, population density, the Gini coefficient, and Moran coefficient, respectively. The Gini coefficient describes the evenness of the distribution of employment or residential areas in a metropolitan area. On the other hand, Moran's coefficient measures the degree of clustering based on employment or residential areas. These metrics provide valuable insights into the general characteristics of an urban area that are crucial for the design of shared mobility modes. For example, research finds that high capacity, high frequency fixed transit is more effective in highly clustered cities compared to a sprawled-out city (Cervero and Seskin, 1995; Dill et al., 2013; Guerra et al., 2018; McIntosh et al., 2014; Stead and Marshall, 2001). Even though the structure of a transportation network is highly correlated with urban form, the metrics in Tsai (2005) may not be sufficient to calculate shareability or provide detailed insights into the design of shared mobility networks like the metrics in the current paper.

Understanding the relationship between urban form, specifically the built environment, and travel behavior is an important and continuing area of research. Ewing & Cervero (2010) present a meta-analysis that analyzes the effects of network and built environment variables on travel behavior; they highlight the variables that are strongly associated with the usage of modes such as personal car, walking, and transit. Oke et al. (2019) present a novel urban typology that spans over 300 cities from across more than 120 countries around the world. The authors use clustering and factor analysis to identify 9 factors and 12 different urban typologies based on the nature of travel behavior and the characteristics of the transportation system. Oke et al. (2019) provide an extensive analysis at an aggregate level in a city to plan sustainable transportation policies based on the urban typology of the city.

Further studies characterize mobility patterns within cities using real trip data and/or transportation networks. Yang et al. (2016) characterize urban spatial-temporal dynamics using cell phone location data for the city of Shenzhen, China. The study identifies 'hotspots' of human convergence and divergence in a city from cell phone data that the authors propose could be used for traffic management and infrastructure planning.

Schieber et al. (2017) evaluate the topography of a network and develop an algorithm to characterize and compare different network structures; however, they do not account for differences between two networks with the same structure but different flow patterns. The current study jointly characterizes the network structure as well as the flow or travel patterns in the network. Batty (2012) states that "to understand cities and their design, they need to be assessed as systems composed of flows and networks, and not as artifacts."

Saberi et al. (2017) present characterize mobility patterns as complex Origin-Destination demand networks to develop aggregate network level statistical properties. They represent trip patterns in a network as a weighted graph of nodes and edges, with nodes representing activity locations (aggregated at a Census Block level) and weights on edges representing the number of person-trips between each pair of nodes. Their metrics include node degree and its coefficient of variation that represent connectivity between activity locations and heterogeneity of connectivity across the network, respectively. They also propose node flux, link weight, network diameter, average shortest path (to represent interaction strength between places); and node and network clustering coefficients (to represent local and aggregate strength of interaction between places in the network). A takeaway from the study is that cities with different urban structures and topologies, such as Chicago and Melbourne, may still exhibit similarities in some aggregate travel characteristics.

## 2.2 Shared Mobility

Despite the potential and initial promise of app-based shared mobility services in terms of making efficient use of space in urban areas, the actual results have been quite mixed or mostly negative thus far. Recent studies have shown that ride-sourcing services—services where each vehicle serves a single traveler request at a time—account for a significant proportion of trips operated by TNCs (Schaller, 2018). These services have been found to be partially responsible for increasing congestion in cities (Erhardt et al., 2019), with 70% of all ridesourcing trips occurring in the 9 most densely populated metropolitan areas in USA (Schaller, 2018). Even shared-ride and ride-splitting services such as Uber Pool, Uber Express Pool, and Lyft Shared Rides have been found to increase net driving on city streets by as much as 160%, since most of the trips are 'new' automobile trips that replaced transit trips or active transportation modes such as walking and biking (Schaller, 2018). Moreover, flexible transit services such as Bridj and Chariot have shut down operations due to low ridership despite offering potential to reshape transit in cities and suburban areas.

TNCs have also been found to have an adverse impact on transit ridership in dense urban centers (Diao et al., 2021; Graehler et al., 2018; Schaller, 2018). Graehler et al. (2019) observes that except for a few years in certain cities, bus transit ridership has declined more steeply compared to rail modes over the past 10 years. The reason is likely that fixed route transit services operating in suburban areas (and even some urban areas) are unable to provide automobile competitive travel times.

These findings suggest that, if app-based mobility services are to improve mobility and accessibility in a sustainable and space-efficient manner, cities are going to need to regulate these services more effectively (Dandl et al., 2021) as well as reconfigure transit systems and design them jointly with emerging app-based mobility services in mind (Liu et al., 2019; Pinto et al., 2019). The authors of the current study hypothesize that the models and shareability metrics proposed herein can support these efforts—a hypothesis we plan to test in subsequent research.

Several studies have explored the behavioral parameters contributing to willingness to share trips and use shared mobility services across different markets and socio-economic groups (Alonso-González et al., 2020; Kang et al., 2021; Shaheen and Caicedo, 2021). Travel behavior can have a major influence on shareability potential in a region. While this study does not directly incorporate travel behavior into the shareability metrics, with the requisite data it is relatively straightforward to include willingness-to-share.

## 2.3 Measuring Shareability in Transportation Networks

Tsao et al. (1999) develop a methodology to analyze a form of sharing potential for carpooling, called the Carpool Demand Reduction (CDR), using the gravity model for a hypothetical area with trips and jobs uniformly distributed throughout the plane. The model assumes a maximum of two passengers involved in a carpool, and that carpooling passengers need to be either from the same origin zone or heading to the same destination zone. The study finds that there is limited potential to reduce demand through carpooling for a region with uniformly dispersed trips and jobs.

Cici et al. (2013) measure sharing potential as the potential reduction in single vehicle trips when travelers are willing to share rides, using Call Description Records to infer home and work locations of travelers in the city of Madrid, Spain. Using the same sharing potential metric as Cici et al. (2013), Gurumurthy & Kockelman (2018) use cell phone traces to infer OD locations and find the potential for dynamic ride sharing in Orlando, Florida. Gurumurthy & Kockelman (2018) employ a dynamic ride-sharing algorithm within an agent-based dynamic simulation model to serve the trips in the cell phone data.

Santi et al. (2014) developed the notion of 'shareability networks' to model trip sharing in a deterministic setting. They apply methods from graph theory to solve the taxi trip sharing problem on the New York City taxi data. Their model includes two parameters—a shareability parameter that determines the number of trips that can be shared by a single vehicle, and the quality of service or detour parameter that defines the total willingness of passengers to detour from their preferred paths. Naturally, as the shareability parameter and the total willingness of passengers to detour increases, the amount of sharing increases and the required vehicles to serve the trips decreases. The main difference between the current study and Santi et al. (2014) is that the latter study focuses on a specific mode with specific vehicle capacities, whereas the current study is mode- and vehicle- agnostic.

Alonso-Mora et al. (2017) extend the shareability network concept developed by Santi et al. (2014) and used it in a dynamic rideshare setting with a capacity of up to 10 passengers per vehicle. The study shows that a significant reduction in single passenger taxi trips in New York City is possible if passengers are willing to take a small detour and share rides with strangers. Tachet et al. (2017) also extend the shareability network notion developed by Santi et al. (2014) and use it to plot shareability curves for 4 cities, namely New York, Vienna, Singapore and San Francisco. The study plots the probability of sharing trips against the trip density, in terms of trips/hour/sq. km. The results show that the curves follow a similar pattern in all the cities, rising steadily at lower densities and quickly reaching saturation.

A report by FHWA defines a 'Shared TNC Overlap Rate' parameter to capture the percentage of a shared trip in which two travelers are in the vehicle together (Middleton et al., 2021). However, this is used only as an input parameter for running different scenarios and not as a metric of shareability for a given network and travel flows.

Kucharski and Cats (2020) develop several person and vehicle level shareability metrics based on their Exact Matching of Attractive Shared rides (ExMAS) algorithm. Unlike the previous studies cited in this sub-section, Kucharski and Cats (2020) analyze trip patterns across all modes in the study region of Amsterdam to evaluate shareability purely from a demand perspective; they match shareable trips to each other rather than to a fleet of vehicles. Kucharski and Cats (2020) measure shareability across different scenarios by varying system parameters such as total demand, matching time window horizon, and price discount for sharing rides compared to ride hailing. Soza-Parra et al. (2021) apply the approach from Kucharski and Cats (2020) to examine shareability by varying spatial distributions of trip patterns for different number of urban centers, destinations, and trip length distributions. Unlike these two studies, the current study models travel demand as controllable flows through a road network, to measure shareability.

## 2.4  Research Gaps Addressed

The current study addresses the following gaps in the research literature. First, this study jointly characterizes the transportation network and the demand that flows through the network to measure shareability. Studies in Section 2.1 either quantify the urban form and land use alone (Srinivasan, 2002; Tsai, 2005); the transportation network alone (Schieber et al., 2017); or travel patterns alone without considering the underlying transportation network (Yang et al., 2016). Even though Saberi et al. (2017) formulate network-based metrics that make use of shortest path distances computed from the underlying transportation network, they do not focus on shareability in their study and therefore do not capture how trip flows between multiple OD pairs in a region can interact within the network to increase overlap and shareability.

Second, except for Kucharski and Cats (2020) and Soza-Parra et al. (2021), the models used in the existing literature to explore shareability (Alonso-Mora et al., 2017; Santi et al., 2014; Tachet et al., 2017) are specific to a single mode—a fleet of shared-ride vehicles in nearly all cases. While these studies and their associated metrics are useful for a specific mobility service, they are less informative than the metrics proposed in the current paper for the design of a future integrated multi-modal shared mobility networks. The current study considers all individual person-trips in any network, irrespective of mode, to holistically capture the shareability of the region.

Finally, the current study formulates a mathematical program to measure shareability agnostic of mode and independent of vehicle type/capacity/fleet-size/routing algorithm, from a *strategic planning* perspective. The objective of the math program is to maximize network flow overlaps by assigning demand from OD pairs to network paths. As far as we are aware, this type of network flow optimization problem is completely novel and may have applications in transportation and other network domains.

## 3    Conceptual Framework

### 3.1    Unit Flow Overlaps

This study introduces the notion of flow overlap as a fundamental unit to measure shareability in a network. Broadly defined, flow overlap (or trip overlap) is the extent of spatial, and temporal in some cases, commonality between the paths of trips in a (transportation) network. Flow overlaps are measured considering all person-trips in the network simultaneously (i.e., total travel demand in the network, irrespective of mode choice) in order to characterize the potential commonalities between OD person-trips in the road network.

As a starting point, let the flow overlap for a unit flow (i.e., a trip) from origin $o$ to destination $d$ on path $k$ be defined as the link-length-weighted average number of other trips/flows—traveling between all OD pairs—with whom the unit flow overlaps on its path $k$. Equation 1 displays the mathematical formulation of flow overlap for a unit flow.

$$Z_{od}^k = \sum_{\forall a \in A} \frac{f_a^{odk} c_a \delta_a^{odk}}{\sum_{\forall a \in A} (c_a \delta_a^{odk})} + (F_{od} - 1) \tag{1}$$

where, $A$ is the set of links in the network, indexed by $a \in A$; $O$ and $D$ are the set of all origins and destinations in the network, indexed by $o \in O$ and $d \in D$, respectively; $K_{od}$ represents a finite set of acyclic paths for OD pair $(o, d)$, indexed by $k \in K_{od}$; $c_a$ gives the length of link $a$; $\delta_a^{odk}$ is a binary parameter that denotes the link-path incidence of link $a$ on path $k$ from OD pair $(o, d)$; $f_a^{odk}$ provides the number of other person-trips with ODs distinct from $o, d$ with whom the original person-trip on path $k$ traverses link $a$; $F_{od}$ represents the number of person-trips heading from $o$ to $d$; and $Z_{od}^k$ represents the average (weighted by link length) trip overlap for a single person-trip on path $k$ between OD pair $(o, d)$.

In Eqn.1, the first term represents the link-length-weighted average number of person-trips traversing the links on path $k$, from OD pairs other than $(o, d)$, with whom a unit flow on path $k$ between OD pair $(o, d)$ shares its path. The denominator of the first term denotes the total length of path $k$ from $o$ to $d$. The numerator weighs the number of all other person-trips not coming from $(o, d)$ that are on link $a$, where the weight is the length of link $a$. The second term in Eqn. 1 denotes the number of other person-trips from the same OD pair $(o, d)$ as the unit person-trip on path $k$. The minus one in the second term represents the unit person-trip on path $k$ for OD pair $(o, d)$ for whom trip overlap is being calculated.

Eqn. 1 can be extended to include a time dimension, $t$, to capture spatial-temporal overlaps during a specific period. Including the temporal component captures the differences in travel patterns and therefore shareability between peak and non-peak hours, or even between AM and PM peak periods. Equation 2 displays the spatial-temporal overlap for a unit flow between origin $o$ and destination $d$ at time $t$.

$$Z_{od}^{kt} = \sum_{\forall a \in A} \frac{f_a^{odkt} c_a^t \delta_a^{odkt}}{\sum_{\forall a \in A} (c_a^t \delta_a^{odkt})} + (F_{od}^t - 1) \qquad (2)$$

where $t$ represents the time interval for which overlap is measured.

Eqn. 1 shows the trip overlap for a unit trip between an OD pair in a transportation network, wherein $f_a^{odk}$ is given. However, ultimately, $f_a^{odk}$ depends on the network path assignment of flows between every other OD pair in the network. As the paths between OD pairs change, so too does $f_a^{odk}$. Hence, to determine the maximum overlap (i.e., shareability) in the network, it is necessary to determine all paths and flows from all OD pairs in the network simultaneously. Moreover, there is inherently a trade-off between maximizing overlaps and minimizing detours when assigning OD flows to paths in a network. After presenting an illustrative example in the next subsection, the following subsection discusses the need for a mathematical formulation to determine optimal paths between all OD pairs that maximizes overlaps in the whole network while considering detours.

## 3.2 Illustrative Example

In this study, the base scenario is the case where all OD flows travel along their shortest paths. Using the formula in Eqn.1, it is straightforward to calculate the flow overlap for every unit flow in the network and then sum all these unit flow overlaps to obtain total and average flow overlaps in the network.

Figure 3-1 and Figure 3-2 illustrate the notion of overlaps in a simple network and show how changes in path assignment can lead to higher flow overlaps. Consider the undirected network with three origin nodes (A, B and C) and one destination node (D), with link distances as shown next to each link in Figure 3-1 and Figure 3-2. Consider the case where three person-trips, X, Y, and Z start their trips from origin nodes A, B, and C, respectively, and all have D as their destination node.

The shortest paths between each OD pair are distinct and do not share any common links. Hence, assigning each of these person-trips to their shortest paths, as shown in Figure 3-1, produces zero trip overlap for each individual trip.

In the second scenario, depicted in Figure 3-2, person-trips X and Z detour from their shortest paths and instead travel to destination node D via node B and link B-D. Person-trip Y continues to take their shortest path to D. This path assignment of person-trips produces flow overlap on link B-D. Person Y shares their trip with 2 others on the entirety of their trip from B to D. Person X and Z do not have any overlap on link AB and CB, respectively, which corresponds to 3/7th of their total trip length. However, from B to D (4/7th of their trip length), X and Z share this link with two other trips.

Using Eqn. 1, the cumulative value of flow overlaps across the three person-trips in Figure 3-1 and Figure 3-2. are 0 and 1.43, respectively. Moreover, while the detour in Figure 3-1 is 0 units, the average detour per person-trip in Figure 3-2 is 1.33 units. Hence, in this simple case, by adjusting the paths of person-trips away from their shortest paths, it is possible to increase overlaps in the network with minor increases in detours.

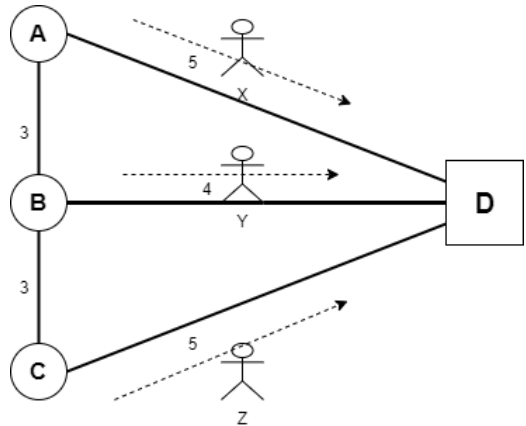

**Figure 3-1 Shortest Path Flows**

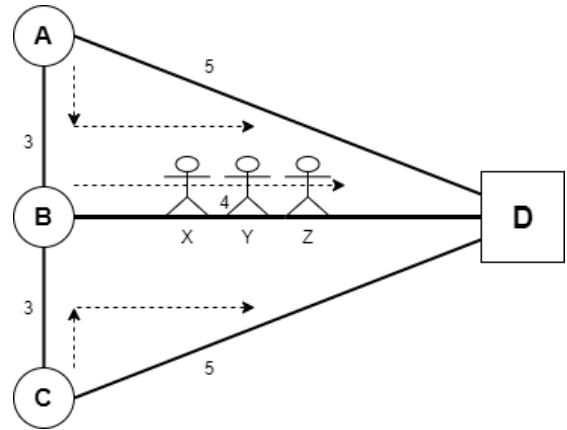

**Figure 3-2 Maximum Overlap Flows**

Although beyond the scope of this study, the path choice depicted in Figure 3-2 makes it possible for the three person-trips to potentially share the same shared mobility mode for a section of their respective journeys. If all three person-trips share the same vehicle on link B-D, it may be possible to substitute three single occupancy vehicle trips on the link with a single share-ride vehicle trip. In this case, the total VMT in the network would decrease from 14 miles in Figure 3-1 to 10 miles in Figure 3-2. However, we want to emphasize that Eqn. 1, Figure 3-1, Figure 3-2, and this paper as a whole, do not consider specific modes or even vehicle trips. The shareability metrics, including flow overlaps, capture person-trips rather than vehicle trips. This paper focuses on person-trips as this study aims to provide insights for strategic planning and multi-shared-mobility-mode network design, rather than for operational planning.

### 3.3 Extending Unit Flow Overlaps to Network Flow Overlap

While it is straightforward to determine the paths between each OD pair that maximize network flow overlap in the illustrative example, the problem is considerably more complex for larger networks, more OD demand pairs, and more paths between each OD pair. Hence, this study proposes a mathematical programming formulation, called the MNFLOP, to determine the optimal path between each OD pair in the network that maximizes cumulative network flow overlap.

Section 4 formulates the MNFLOP as a flow- and path-based quadratic integer program. The proposed MNFLOP model quantifies shareability in a transportation network or subnetwork, given two inputs: a road network, and origin-destination (OD) person demand flows. The MNFLOP model outputs, or decision variables, include the flow on each network link and the single path for each OD pair.

Given the assigned path for each OD pair, it is possible to calculate many other shareability metrics in addition to the MNFLOP objective function term(s). The MNFLOP objective function term derived from the unit flow metric in Eqn. 1, is the primary metric for characterizing the shareability of an urban area. However, Section 5 provides a plethora of other shareability metrics obtainable from the output of the MNFLOP model.

# 4 Mathematical Formulation

## 4.1 Problem Statement

For a given network and OD flow-based demand, the objective of the MNFLOP is to maximize total flow overlap in the network via assigning demand between every OD pair to a network path $k$. Since the MNFLOP involves assigning trips to non-shortest paths, the objective function includes a second term to account for tradeoffs between overlaps and detours. Equation 3 provides a formulation of the MNFLOP objective function considering the flows between all OD pairs in the network.

$$Objective = Max\, Z = \sum_{\forall o \in O} \sum_{\forall d \in D} F_{od} \sum_{\forall k \in K_{od}} X_{od}^k \left( Z_{od}^k - w\Delta_{od}^k \right) \tag{3}$$

All variables in Eqn. 3 are the same as defined previously. The new parameters, indices, and decision variables are as follows: $X_{od}^k$ is a binary decision variable equal to 1 if the flow for OD pair $(o, d)$ is assigned to path $k$; $\Delta_{od}^k$ represents total detour on path $k$ compared to the shortest path for OD pair $(o, d)$; and $w$ is a weight parameter that effectively converts units of detour distance into units of overlap. A higher value for detour parameter $w$ favors the assignment of trips on their shortest/preferred paths.

## 4.2 Problem Formulation

Combining Eqn. 1 and Eqn. 3, Eqn. 4 displays the objective of the bi-criteria MNFLOP that incorporates all the key parameters as well as the auxiliary decision variables, $f_a^{odk}$:

$$Max\, Z = \sum_{\forall o \in O} \sum_{\forall d \in D} F_{od} \left[ \sum_{\forall k \in K_{od}} X_{od}^k \left( \sum_{\forall a \in A} \frac{f_a^{odk} c_a \delta_a^{odk}}{\sum_{\forall a \in A}(c_a \delta_a^{odk})} + (F_{od} - 1) - w\Delta_{od}^k \right) \right] \tag{4}$$

All the indices, parameters, and decision variables in Eqn. 4 are as defined above.

Eqn. 4 includes $f_a^{odk}$, a set of auxiliary decision variables, denoting the number of overlapping trips on link $a$ of path $k$ from OD pair $(o, d)$ from OD pairs other than OD pair $(o, d)$. Equation 5 displays to formula for these auxiliary decision variables.

$$f_a^{odk} = \sum_{\forall (o',d') \in (O \times D - \{(o,d)\})} F_{o'd'} * \sum_{k' \in K_{o'd'}} X_{o'd'}^{k'} * \delta_a^{o'd'k'} * \gamma_{o'd'k'}^{odk} \tag{5}$$

where, $(o', d')$ denotes any OD pair different from $(o, d)$ and all the other decision variables and parameters are as defined above. The new parameter, $\gamma_{o'd'k'}^{odk}$ denotes an adjustment factor (described further in section 4.2.1) that aims to prevent choosing paths with unnecessary detours for a pair of ODs.

The auxiliary decision variable $f_a^{odk}$ is computed for all links $a$ that belong to path $k$ from all OD pairs in the network. Since the path assignment for maximum trip overlap of an OD pair depends on the path assignment of many other OD pairs, this implies that the MNFLOP is a Quadratic Integer Programming Problem.

The MNFLOP is subject to the constraint that all trips between an OD pair must be assigned to one and only one path, stemming from Eqn. 6 and the binary decision variable constraint.

$$\sum_{\forall o \in O} \sum_{\forall d \in D} \sum_{\forall k \in K_{od}} X_{od}^k = 1 \quad \forall (o,d) \in O \times D \tag{6}$$

$$X_{od}^k \in 0,1 \quad \forall o \in O, d \in D, k \in K_{od}$$

### 4.2.1 The auxiliary flow adjustment factor $\gamma$

As described in the previous sections, MNFLOP tries to maximize flow overlap in the network, which is the link-length-weighted average of overlapping flows on the chosen path between each OD pair (as shown in Eqn. 4). One issue with this objective is that the solver will look to assign overlapping flows from two or more ODs to the longest common overlapping path possible. Assigning flows from a pair of ODs to longer common paths may not even increase the number of overlapping flows; it may just force the overlapping flows to travel longer distances together.

Figure 4-1 displays such an example. The network on the left of Figure 4-1 shows the result of MNFLOP for two OD pairs, (2,16) and (6,16), wherein the 'optimal' paths for each OD pair include a detour onto sub-path 8-7-18-16, rather than just traveling directly from 8 to 16, despite the detour not resulting in any additional flow overlaps.

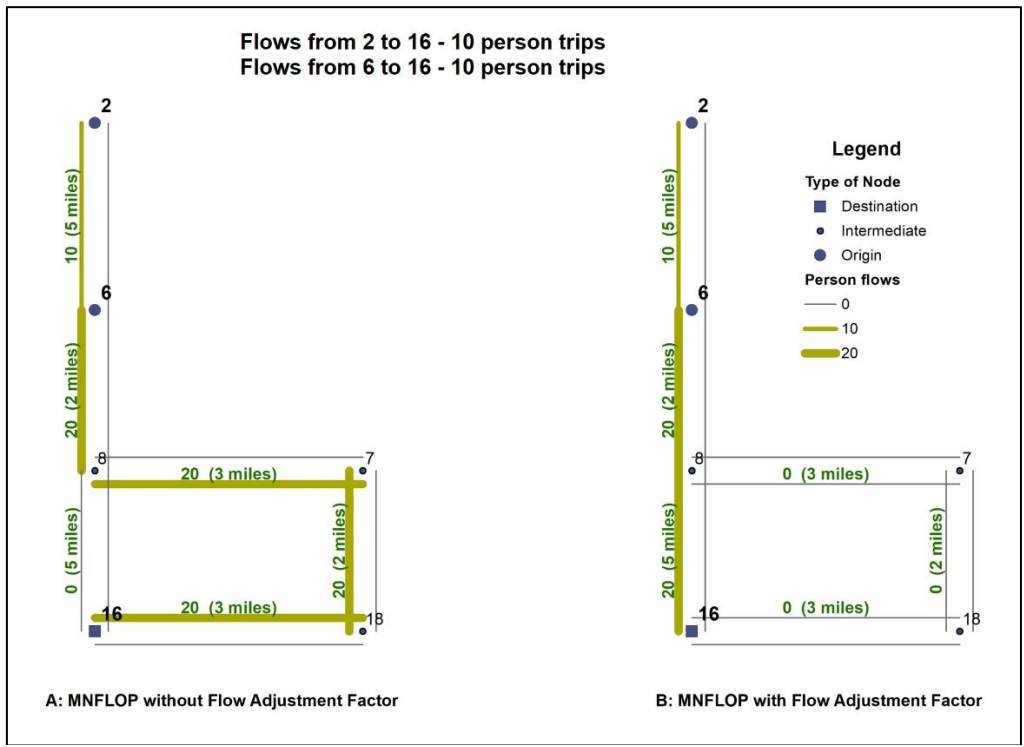

**Figure 4-1 MNFLOP Path Choices with and without Auxiliary Flow Adjustment Factor**

The average flow overlap for a unit flow from 2 to 16 ($Z_{2,16}^{k*}$) is 15.66 person-trips on the left side and 14.83 person-trips on the right side of Figure 4-1, respectively. This gain of 0.83 overlapping person-trips is because a unit flow from 2 to 16 on the left map shares a longer path with 10 overlapping flows from 6 to 16 for $\left(\frac{10}{15}\right)^{\text{th}}$ of its total path ($10 * \frac{10}{15} = 6.66$ person-trips) compared with the right map where a unit flow shares the shortest path with the same 10 overlapping flows for $\left(\frac{7}{12}\right)^{\text{th}}$ of its total path ($10 * \frac{7}{12} = 5.83$ person-trips). Therefore, MNFLOP assigns trips to the longer path with a detour of three miles just to maximize overlap distance without actually gaining any additional overlapping flows.

To prevent this undesirable outcome, we include an 'auxiliary flow adjustment factor', $\gamma_{o'd'k'}^{odk}$, that discounts overlapping flows from every 'other' OD pair $(o', d')$ when calculating the auxiliary decision variable for path $k$ between OD pair $o, d$ $(f_a^{odk})$.

The following steps describe the procedure to discount flows for all combinations of paths between two pairs of ODs, $(o, d)$ and $(o', d')$, using the auxiliary flow adjustment factor, $\gamma_{o'd'k'}^{odk}$.

1. Find the pair of paths, $(k_*, k_*') \in K_{OD} \times K_{O'D'}$, that have the shortest distance based on summing (i) the distance of the overlapping sub-path between path $k$ and path $k'$, (ii) the distance of the non-overlapping sub-path for path $k$, and (iii) the distance of the non-overlapping sub-path for path $k'$. Equation 7 displays the mathematical formulation for this step.

$$(k_*, k_*') = argMin_{k,k'} \left( \sum_{a_1 \in A_{kk'}} c_{a_1} + \sum_{\forall a_2 \in A_k} c_{a_2} + \sum_{a_3 \in A_{k'}} c_{a_3} \right) \tag{7}$$

where,

$A_{kk'}$ is the set of links common to both paths $k$ and $k'$.

$$A_{kk'} = \{a \in A : \delta_a^{odk} = 1\} \bigcap \{a \in A : \delta_a^{o'd'k'} = 1\} \tag{8}$$

$A_k$ and $A_{k'}$ correspond to the set of links on paths $k$ and $k'$ respectively that are not shared.

$$A_k = \{a \in A : \delta_a^{odk} = 1\} \setminus A_{kk'} \tag{9}$$

$$A_{k'} = \{a \in A : \delta_a^{o'd'k'} = 1\} \setminus A_{kk'} \tag{10}$$

$O'D'$ represents the set of all OD pairs other than $(o, d)$ for which overlap is being measured.

$$O'D' = O \times D \setminus \{(o, d)\} \tag{11}$$

In the example shown in Figure 4-1, $(o, d) = (2,16)$ and $(o', d') = (6,16)$. Eqn. 7 returns the shortest paths for both OD pairs, $(k_*, k_*') = (0,0)$, which corresponds to the map on the right of Figure 4-1. In this case, the value of $\left( \sum_{a_1 \in A_{kk'}} c_{a_1} + \sum_{\forall a_2 \in A_k} c_{a_2} + \sum_{a_3 \in A_{k'}} c_{a_3} \right) = (7 + 5 + 0)$ is 12 miles.

2. Calculate the ratio of (i) the length of the overlapping sub-path for the optimal path pair $(k_*, k_*')$, over (ii) the length of path $k_*$.

$$\beta_{o'd'k_*'}^{odk_*} = \frac{\sum_{\forall a \in A_{k_* k_*'}} c_a}{\sum_{\forall a \in A_{k_*}} c_a} \tag{12}$$

This is the optimal overlapping path length fraction for OD pair combination $(o, d)$ and $(o', d')$ when calculating auxiliary flows for a unit flow between OD pair $(o, d)$. This value is needed to discount potential flows on non-optimal path $k'$ from OD pair $(o', d')$ when determining the path $k$ for OD pair $(o, d)$ and calculating auxiliary flows. In the example shown in Figure 4-1, $\beta_{6,16,k'_*}^{2,16,k_*} = \frac{7}{12}$.

3. Calculate the auxiliary flow adjustment factor, $\gamma_{o'd'k'}^{odk}$, for the $(o, d, k, o', d', k')$ tuple. For the optimal combination $(k_*, k'_*) \in K_{OD} \times K_{O'D'}$, an adjustment to the auxiliary flow is unnecessary, therefore, $\gamma_{o'd'k'}^{odk} = 1$. Flow adjustment is also unnecessary when the optimal path combination $(k_*, k'_*)$ has no overlapping links $\left(\beta_{o'd'k'_*}^{odk_*} = 0\right)$. However, for other $k \in K_{OD} \backslash k_*$, Eqn. 13 displays the formula for the auxiliary flow adjustment factor to adjust overlapping flows from $(o', d')$ on path $k'$ while measuring overlap for a unit flow between OD pair $(o, d)$ on path $k$.

$$\gamma_{o'd'k'}^{odk} = \begin{cases} \dfrac{\beta_{o'd'k'_*}^{odk_*}}{\beta_{o'd'k'}^{odk}} & if \ \beta_{o'd'k'}^{odk} > \beta_{o'd'k'_*}^{odk_*} > 0 \\ 1 & otherwise \end{cases} \tag{13}$$

where $\beta_{o'd'k'}^{odk}$ is the fraction of length that path $k$ from OD pair $(o, d)$ overlaps with path $k'$ from OD pair $(o', d')$, and $\beta_{o'd'k'_*}^{odk_*}$ is the optimal overlapping path length fraction (from step 2) for the tuple $(o, d, o', d')$.

The auxiliary flow adjustment factor ranges between 0-1 and ensures that overlapping flows from other OD pairs $(o', d')$ do not force unnecessary detours for flows from OD pair $(o, d)$.

In the example shown on the left in Figure 4-1, where flows from 2 to 16 and 6 to 16 are assigned to the longer common path $(k, k') = (1,1)$ via nodes 7, 18, and 16, the adjustment factor value is $\gamma_{6,16,1}^{2,16,1} = \frac{\frac{7}{12}}{\frac{10}{15}} = \frac{7}{8}$. As demonstrated earlier in this section, overlapping flows from 6 to 16 while measuring overlap for a unit flow from 2 to 16 are inflated (6.66 person-trips) in the left of Figure 4-1. Multiplying these flows by the auxiliary flow adjustment factor $\gamma_{6,16,1}^{2,16,1} = \frac{7}{8}$ in Eqn. 5 ensures that overlapping flows from 6 to 16 are no longer inflated in the auxiliary flow calculation in Eqn. 4. The $\frac{7}{8}$ adjustment factor reduces the auxiliary flow from OD pair $(o', d') = (6,16)$ on path $k' = 1$ to $(10 * \frac{10}{15} * \frac{7}{8}) = 5.83$ person-trips, when calculating auxiliary flow for OD pair $(o, d) = (2,16)$. This ensures that MNFLOP only assigns flows from 2 to 16 and 6 to 16 on longer paths if the longer path increases overlapping flows from other OD pairs.

Steps 1 and 2 are repeated for every combination of $(o, d)$ and $(o', d')$ pairs in the network. The auxiliary flow adjustment factor, $\gamma_{o'd'k'}^{odk}$ in step 3 is calculated for all path combinations $(k, k')$ for each pair of $(o, d)$ and $(o', d')$ combinations in the network.

## 4.3 Computational Complexity Considerations

The section provides a brief overview of the computational complexity associated with the MNFLOP. If a network has $|O|$ origins, $|D|$ destinations, and $|K|$ paths between each OD pair, then the number of path choice decision variables in the optimization problem is $|O| \times |D| \times |K|$. The number of path choice decision variables is independent of the number of links in the network and depends only on the number of OD pairs and the paths between them. However, the quadratic nature of the objective function in Eqn. 4 means that an optimal path choice for an OD pair is strongly influenced by the optimal path choice for all other OD pairs with which there is an overlapping link on their paths. The number of quadratic terms in the objective function increases as the OD pairs increase in spatial closeness, as this results in more overlapping links between paths from multiple OD pairs. Naturally, the number of quadratic terms also increases with network size, as larger networks have more links on which different OD pairs are likely to have overlapping paths.

For a network with $|O|$ origins, $|D|$ destinations and $|K|$ paths between each OD pair, the MNFLOP could have up to a maximum of $\binom{|O| \times |D| \times |K|}{2} - (|K| - 1) \times |O| \times |D|$ quadratic terms in the objective function. $\binom{|O| \times |D| \times |K|}{2}$ is the number of combinations of overlapping paths, and $(|K| - 1) \times |O| \times |D|$ is the number of other paths between the same OD pair. More quadratic terms results in slower convergence while trying to find an exact optimal solution.

## 4.4 Solution Methodology

The MNFLOP is solved as a constrained Mixed Integer Quadratic Problem (MIQP) using the Gurobi optimization package in Python programming language. The Networkx package is used to create the network and perform other network related operations. As an input to the MNFLOP a finite set of $|K|$ acyclic shortest distance paths are pre-computed for each OD pair: $K_{od} = 0, 1, 2, \ldots |K|$ indexed by $k \in K_{OD}$, where $k = 0$ denotes the shortest path between OD pair $(o, d)$, $k = 1$ denotes the second shortest path and so on. Yen's algorithm finds the $k$ shortest paths (Yen, 1971), where $k$ is set to 5 in this study. In practice, there are many ways to determine $K_{od}$ and the size of $K_{od}$, such as including all paths between OD pair $(o, d)$ with a detour less than $x$ miles and/or less than $y$ percent longer than the shortest path.

## 5 Shareability Metrics

The output of the MNFLOP model can provide a variety of shareability metrics at different levels of aggregation, such as the OD level, link level, node level (origin or destination), and the network level. Table 5.1 provides a comprehensive list, along with descriptions and formulations, of shareability metrics.

Since flow overlap, the measure of shareability in this study, is defined at the OD level (for a unit trip), most of the OD-level metrics in Table 5.1 come directly from the conceptual framework section. Trip overlap, $Z_{OD}^k$ in Eqn. 1, is the fundamental unit of shareability; trip overlap percentage normalizes flow overlap for an OD pair by all OD demand; detour distance for a path $k$ between an OD pair, $\Delta_{OD}^k$, is straightforward and measured relative to the shortest path distance; the marginal overlap, $M_{Z_{od}^k}$, captures the ratio of overlap increase to detour distance when switching from the shortest path to path $k$ for an OD; and finally, overlap distance, $L_{od}^k$, parallels trip overlap but instead of dividing the sum product of link flows and link distances by total path distance, overlap distance divides the sum product by total OD demand.

The node-level and network-level metrics parallel the OD-level metrics. In fact, all the node-level and nearly all the network-level metrics can be derived by summing the OD-level \metrics.

**Table 5.1 Shareability Metrics**

| Metric | Unit | Aggregation Level | Description | Formula |
|---|---|---|---|---|
| *Trip Overlap* | Person-trips | OD | Average number of other flows with which a unit trip from O to D on path $k$ shares its path (Eqn. 1) | $Z_{od}^k$ |
| *Trip Overlap Percentage* | % | OD | Percentage of other flows in the network with which a unit trip from O to D on path $k$ shares its path, where $F$ is total OD flow. | $Z_{od}^{k,\%} = \dfrac{Z_{od}}{F-1}100$ |
| *Detour* | Miles | OD | Difference in distances between the chosen path $k$ and the shortest path from O to D | $\Delta_{od}$ |
| *Marginal Overlap* | Person-trips/mile | OD | Marginal change in overlapping flows for every detour mile obtained from shifting from the Shortest Path to path $k$ for an OD | $M_{Z_{od}^k} = \dfrac{Z_{od} - Z_{od}^{SP}}{\Delta_{od}}$ |
| *Overlap Distance* | Miles | OD | Average distance a unit trip on path $k$ from O to D shares with all other flows in the network | $L_{od}^k = \sum\limits_{\forall a \in A} \dfrac{c_a * \delta_a^{odk} * (f_a^{odk} + F_{od} - 1)}{F-1}$ |
| *Node Overlap* | Person-trips | Origin Node | Average number of other flows from an origin that a unit flow originating from the same node shares its path. | $Z_o = \sum\limits_{\forall d \in D} \dfrac{Z_{od} F_{od}}{F_o}$ |
| *Node Overlap Percentage (Dispersion)* | % | Origin Node | Percentage of demand that originates from a node that share paths in whole or in part with other flows from the same node. | $Z_o^\% = \sum\limits_{\forall d \in D} \dfrac{Z_{od}^\% F_{od}}{F_o}$ |
| *Node Overlap Distance* | Miles | Origin Node | Average distance a person-trip overlaps with all other trips from the same origin. | $L_o = \sum\limits_{\forall d \in D} \dfrac{L_{od} * F_{od}}{F_o}$ |
| *Average Network Overlap* | Person-trips | Network | Average number of other flows with which a unit trip in the network shares its path | $Z = \sum\limits_{\forall o \in O} \sum\limits_{\forall d \in D} \dfrac{Z_{od} F_{od}}{F}$ |
| *Average Network Overlap Percentage* | % | Network | Percentage of total demand in the network that share paths | $Z^\% = \sum\limits_{\forall o \in O} \sum\limits_{\forall d \in D} \dfrac{Z_{od}^\% F_{od}}{F}$ |
| *Average Network Detour* | Miles | Network | Demand weighted average of detour miles for all flows in the network | $\Delta = \sum\limits_{\forall o \in O} \sum\limits_{\forall d \in D} \dfrac{\Delta_{od} F_{od}}{F}$ |
| *Average Network Detour Ratio* | - | Network | Ratio of average trip length in MNFLOP assignment to the average length of trips in Shortest Path assignment | $\Delta_r = \dfrac{Avg\ Trip\ Dist_{MNFLOP}}{Avg\ Trip\ Dist_{SP}}$ |
| *Marginal Network Overlap* | Person-trips/mile | Network | Total increase in overlap in the network using MNFLOP assignment for a unit detour from Shortest Path | $M_Z = \dfrac{Z^{MNFLOP} - Z^{SP}}{\Delta}$ |
| *Marginal Network Overlap Percentage* | % | Network | Total increase in percentage of overlapping flows in the network using MNFLOP assignment for a unit detour from Shortest Path | $M_{Z_\%} = \dfrac{Z_\%^{MNFLOP} - Z_\%^{SP}}{\Delta}$ |
| *Links Used* | Number | Link/Network | Number of links with non-zero person-trip flows | - |
| *Avg Link Flows* | Person-trips | Link/Network | Link-length-weighted-average person-trips on links that have positive flows | - |

At a node level, the metrics of shareability characterize the 'dispersion' of trips originating from or destined to a node. Trips originating from a node exhibit a high degree of dispersion when their trip-ends are scattered to destinations in the network such that the paths to these destinations have limited overlaps. Conversely, when trips from a node are bound to a limited number of destinations and/or the paths to the destinations are highly overlapping, trips from the node exhibit low dispersion. The node overlap percentage metric shown in Table 5.1 quantifies dispersion of trips from a single node. A high value of node overlap percentage indicates that trips from the node are highly concentrated. Additionally, node overlap distance denotes the average distance for which trips originating from a node share paths with other trips originating from the same node. To understand the different between node overlap and node overlap distance, consider two nodes, Node-A and Node-B, that have similar overlap and overlap percentage. If Node-A has a higher node overlap distance, then trips from Node-A share paths for longer distances than trips from Node-B.

At the link level, shareability metrics include the number of person-trips on each link, as well as the number of links in the network that have non-zero flow values. The link flows can also be aggregated at the network level to find the link-length-weighted average number of person-trips on links with non-zero flows. A higher number of person-trips per link with non-zero flows indicates a higher concentration of trips in a region onto fewer links and potentially fewer corridors, the objective of MNFLOP.

## 5.1 Toy Example to Illustrate Node-level Metrics

The node level metrics of shareability presented in this paper consider the magnitude of demand, the spatial distribution of demand (dispersion) as well as the overlapping distance of demand emanating from a location, rather than just measuring demand densities (trips per time unit per area unit).

Figure 5-1 displays four different toy networks to illustrate the calculation of node level measures and their potential value for characterizing shareability from a single node. Notably, all four networks have the same demand density originating from Node-A: 90 trips per time unit per area unit. In scenario 1, all flows from Node-A are bound to Node-B. In scenarios 2 to 4, flows from Node-A are equally distributed among nodes B, C and D. A visual inspection of the scenarios in Figure 5-1 indicates clear differences in flow overlap and dispersion, and therefore differences in shareability across the scenarios.

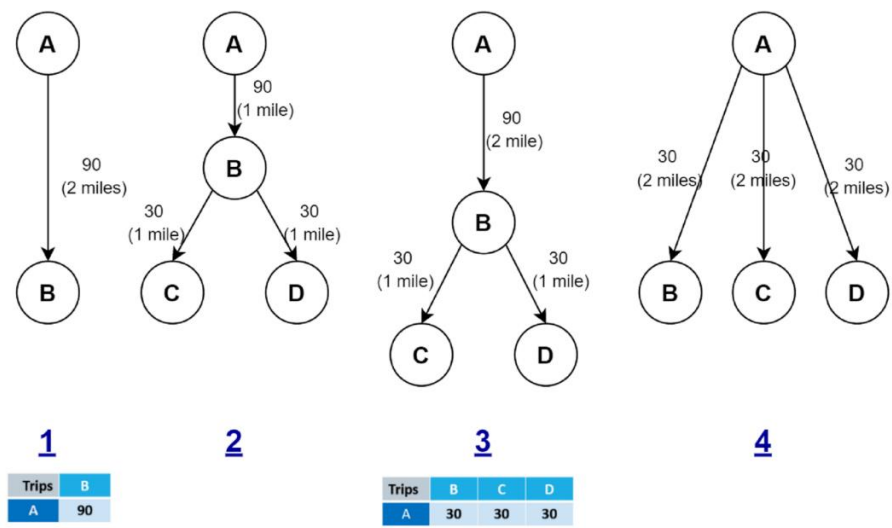

**Figure 5-1 Example Scenarios to Illustrate Overlap/Dispersion and Overlap Distance at Node Level**

The node level overlap metrics in Table 5.2 based on the formulations in Table 5.1 confirm significant differences in the toy network in terms of the various shareability measures. In Scenario 1, all trips are bound to a single destination; hence, all flows share the same path, making the demand emanating from Node-A highly concentrated; the overlap percentage is 100% (i.e., 0 dispersion). A unit trip in this scenario overlaps with 89 other trips. Overlap distance in this scenario is same as the average trip distance since a unit trip from Node-A overlaps with all other flows from Node-A on its entire path.

Scenario 4 exhibits the highest dispersion of trips because a unit trip from Node-A on average shares paths with only 29 other trips (or 32.6% of all trips in the network) for an average overlap distance of 0.65 miles. The average overlap distance in this scenario being much less than the average trip distance indicates that each trip in this network shares paths with only a fraction of all other trips along each trip's respective path.

Scenarios 2 and 3 have nearly similar network structures except for the length of the link connecting Node-A and Node-B. The values of overlap and overlap percentage are slightly higher for scenario 3 than scenario 2 because flows share paths for a longer distance on link AB in scenario 3 than scenario 2. The value of average overlap distance for scenario 3 is one-mile more than scenario 2, which means that the average trip from Node-A in scenario 3 overlaps with all other flows for a mile longer than an average trip from Node-A does in scenario 2.

**Table 5.2 Comparison of Node Level Overlap Metrics for Example Scenarios**

| Metrics of Shareability for Node - 1 | Scenario 1 | Scenario 2 | Scenario 3 | Scenario 4 |
|---|---|---|---|---|
| Total Demand | 90 | 90 | 90 | 90 |
| Avg Overlap (Person-trips) | 89 | 69 | 75.7 | 29 |
| Avg Overlap (%) | 100 | 77.5 | 85 | 32.6 |
| Avg Trip Distance (Miles) | 2 | 1.67 | 2.67 | 2 |
| Avg Overlap Dist. (Miles) | 2 | 1.22 | 2.22 | 0.65 |

# 6 Case Studies

## 6.1 Study Network

To verify the proposed metrics and to illustrate their usefulness for characterizing shareability, this study employs the Sioux Falls road network (Stabler, 2019). The Sioux Falls network includes 24 nodes and 76 directional links (Figure 6-1). The study employs Gurobi's MIQP solver to solve the MNFLOP model. The solver runs on a system with 64 GB RAM and Intel i9 processor with a clock speed of 3.60 GHz. Yen's algorithm is used to generate 5 shortest distance paths between each pair of demand nodes in the network (Yen, 1971). The generated paths are then filtered based on absolute and relative maximum detour distance constraints -- only those paths with detour travel distances less than or equal to the minimum of the absolute and relative maximum detour distances are retained in the set of paths between each OD pair.

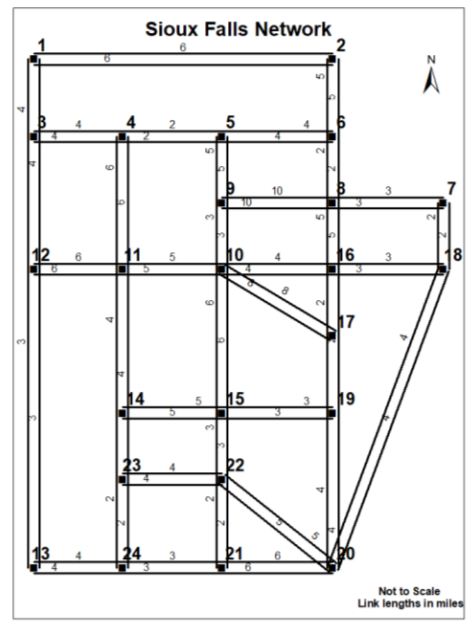

**Figure 6-1 Sioux Falls Street Network**

This study uses an absolute maximum detour distance of 8 miles and a relative maximum detour threshold of 50% of the shortest path travel distance for each OD pair.

## 6.2  Scenarios for Analysis

Based on the objective functions in Eqn. 3 and Eqn. 4, the MNFLOP can be formulated in three different ways, by adjusting the $w$ parameter:

1. **Pure Maximum Network Flow Overlap (P-MNFLOP) Assignment:** Trips between OD pairs are assigned to paths to maximize the total trip overlap in the network. This is achieved by setting $w$ to a small non-zero value in the objective function, such as $w = 0.1$. A small non-zero value is used instead of zero in order to prevent the solver from returning inferior or pareto-inefficient solutions, when considering both overlap and detour. A small value of $w = 0.1 \, overlapping \, trips/mile$ indicates that the minimum marginal overlap gain for a unit trip in the network is 0.1 trips for every detour mile, or about 1 overlapping trip per 10 miles of detour.

2. **Shortest Path (SP) Assignment:** All trips between an OD pair are assigned to the shortest path between their OD pair. This is achieved by setting $w$ to a very high value, close to infinity.

3. **Bi-criterion Maximum Network Flow Overlap (Bi-MNFLOP) Assignment:** This option assigns trips to paths by considering both terms in the objective function. This is achieved by varying the value of $w$ between a small number and a finite large number.

Table 6.1 provides a summary of the various scenarios analyzed in this study. The scenarios vary in terms of Origin and Destination nodes, total demand, and flow assignment methods. Demand is measured as person-trips throughout the results section.

**Table 6.1 Summary of Scenarios for Analysis**

| Scenario | Description | Pure Maximum Overlap (P-MNFLOP) | Shortest Path (SP) | Bi-criterion MNFLOP (Bi-MNFLOP) |
|---|---|---|---|---|
| Scenario 1 | 6 Origin Nodes in North to 6 Destination Nodes in South (Total Demand = 6,000 trips) | Yes | Yes | Yes |
| Scenario Set 2 | Origin Node to all 24 Demand Nodes | Yes | Yes | No |
| Scenario Set 3 | Same Total Demand (1000 trips), Same Set of Origin Nodes, Different Destination Nodes | Yes | No | No |
| Scenario Set 4 | Same Total Demand (1000 trips) Distributed between Different OD pairs | Yes | No | No |

## 7  Results

### 7.1  Scenario Set 1: Comparing Assignment Methods in Baseline Scenario

In this scenario, trips between 6 origin nodes in the North and 6 destination nodes in the South of the Sioux Falls network are assigned using the three methods listed in Section 6.2. A detour weight value of $w = 1800$ trips per mile was used for the baseline Bi-MNFLOP assignment.

Figure 7-1, Figure 7-2, and Figure 7-3 show the link flows on the Sioux Falls network for the P-MNFLOP, SP, and Bi-MNFLOP assignments, respectively. Link thickness is proportional to the number of trips assigned on each link. The plots show that P-MNFLOP concentrates trips between O-D pairs onto the fewest links, followed closely by Bi-MNFLOP, whereas SP spreads person-trips across a large number of links. These plots clearly illustrate the ability of MNFLOP-based assignment to significantly

increase flow overlap compared to SP assignment. The results in Table 7.1 quantify the overlap and detours associated with each assignment method.

Table 7.1 provides the network level shareability metrics for Scenario 1. Many of the metrics directly confirm that P-MNFLOP significantly increases flow overlap, as does Bi-MNFLOP, compared to SP assignment. However, Table 7.1 also indicates that this increase in flow overlap comes at a relatively small cost in terms of detour distance for the Sioux Falls network.

**Table 7.1 Network Level Overlap Metrics, Sioux Falls Scenario-1**

| Metric | P-MNFLOP | SP | Bi-MNFLOP (w=1800) |
|---|---|---|---|
| Avg Overlap (Person-trips) | 2951 | 1725 | 2809 |
| Avg Overlap (%) | 46.1 | 27.0 | 43.9 |
| Avg Trip Distance (Miles) | 16.4 | 15.6 | 16.2 |
| Average Detour (ratio) | 1.05 | - | 1.03 |
| Marginal Overlap (Trips/Detour miles) | 1656 | - | 2007 |
| Marginal Overlap % (Overlap %/Detour mile) | 25.9 | - | 31.4 |
| Avg Overlap Distance (Miles) | 7.3 | 4 | 6.9 |
| Distance Elasticity of Overlap | 15 | - | 18.2 |
| Avg Link Flow (Person-trips) | 1454 | 847 | 1055 |
| # Links with non-zero Flows | 21 | 31 | 26 |

According to Table 7.1, when going from SP to P-MNFLOP, trip overlap in the network increases by nearly 70% (from 1725 to 2950 person-trips) for a mere 5% increase in distance (15.6 to 16.4 miles). This suggests that small detours from shortest paths can significantly increase flow overlaps. As expected, Bi-MNFLOP assignment produces slightly less overlap and lower detours than P-MNFLOP. However, Bi-MNFLOP still significantly increase overlap with only a 3% increase in distance compared to SP.

The marginal overlap values for P-MNFLOP and Bi-MNFLOP show that marginal overlap is higher for Bi-MNFLOP than P-MNFLOP (2007 vs. 1656 trips per detour mile). This stems directly from the detour weight $w = 1800$ trips/mile ensuring a marginal overlap of at least 1800 trips per detour mile.

The average overlap distance for a unit trip in the network increases from 4 miles to 7.3 miles when optimal paths are assigned based on P-MNFLOP compared to SP assignment. This implies that, moving flows away from the shortest path not only increases the number of overlapping trips for a unit flow, but also the distance along the path in which the average person-trip shares path with all other trips.

The average link flows on links with non-zero person-trips also show the highest value for P-MNFLOP, followed by Bi-MNFLOP and SP. This metric along with the number of links with non-zero flows in each case shows that P-MNFLOP effectively concentrates more person-trips onto fewer links.

Figure 7-4 and Figure 7-5 display histograms of trip overlap percentage and trip overlap, respectively, across the 3 assignment methods, where the count variable is number of OD pairs. The overlap and overlap percentage plots show that the histogram for SP assignment skews leftward compared to the other 2 assignment methods. Assigning trips based on P-MNFLOP pushes the distribution rightwards, meaning that the overlap and overlap percentage of most O-D pairs increase relative to SP. For the chosen value of w, the results of the Bi-MNFLOP are closer to P-MNFLOP than the SP.

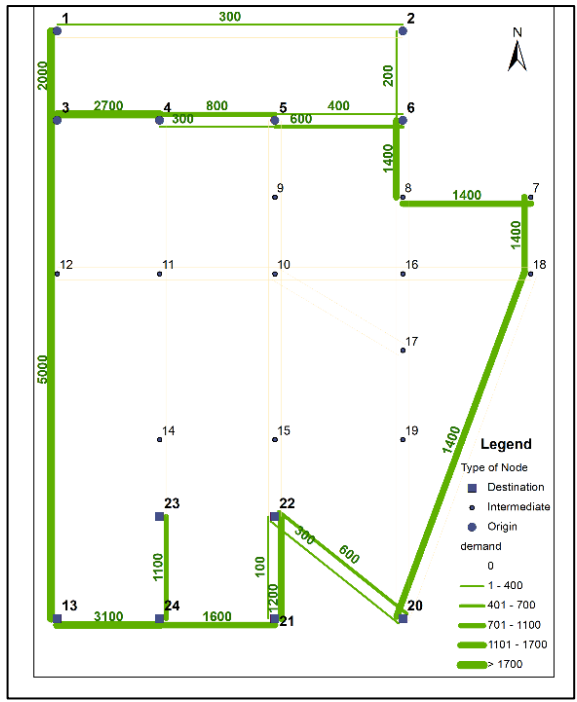

**Figure 7-1 Scenario 1 P-MNFLOP Links Flows**

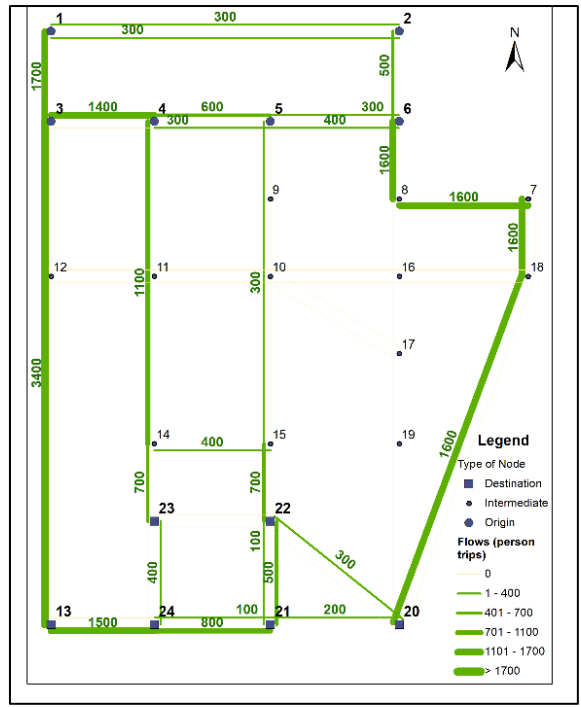

**Figure 7-2 Scenario 1 SP Assignment Links Flows**

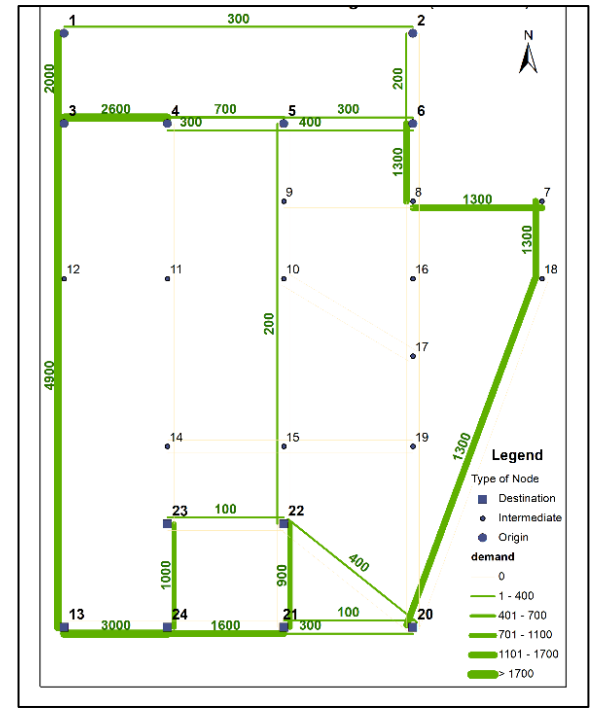

**Figure 7-3 Scenario 1 Bi-MNFLOP Links Flows**

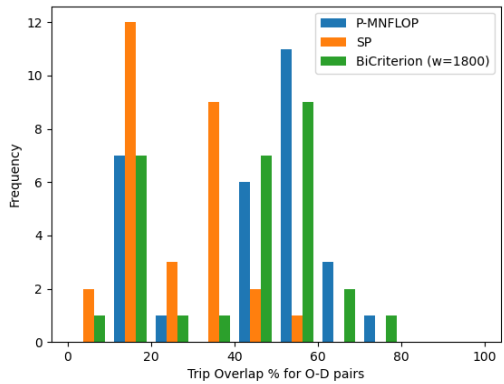

**Figure 7-4 No. of OD pairs and Overlap Percentage**

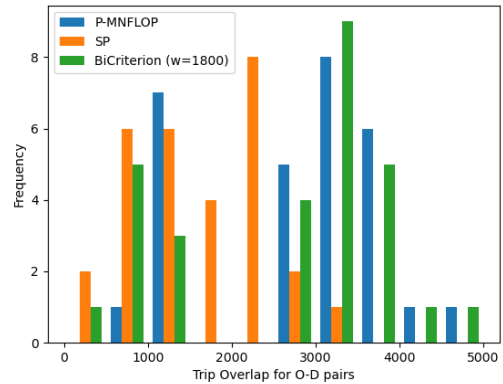

**Figure 7-5 No. of OD pairs and Trip Overlap**

### 7.1.1    Sensitivity Analysis for *w*

The results of Bi-MNFLOP are highly sensitive to *w*. Figure 7-6 displays results for Bi-MNFLOP as *w* ranges between 0.1 and 6,000 trips per detour mile, approximating the Pareto Frontier. A low value of *w* of 0.1 or near 0.1 yields the highest average trip overlap and highest average detour, effectively P-MNFLOP. As expected, as the value of *w* increases, the average trip overlap and average detour decrease. At a *w* value close to 4000, the average detour becomes 0, which means that the Bi-MNFLOP objective has collapsed into an SP assignment method.

As is clear from the plot, changes in *w* do not cause a smooth change in overlap and detour. For example, when *w* ranges from 1800 to 1950 the average detour reduces by 70% ($\Delta$ = 0.54 miles to 0.16 miles), with a similarly large drop in average overlap (25% drop). The lack of smoothness stems from the Boolean nature of the path choice decision variable in the optimization problem, the finite set of paths between ODs (which satisfy maximum detour distance constraints), and the relatively small number of OD pairs.

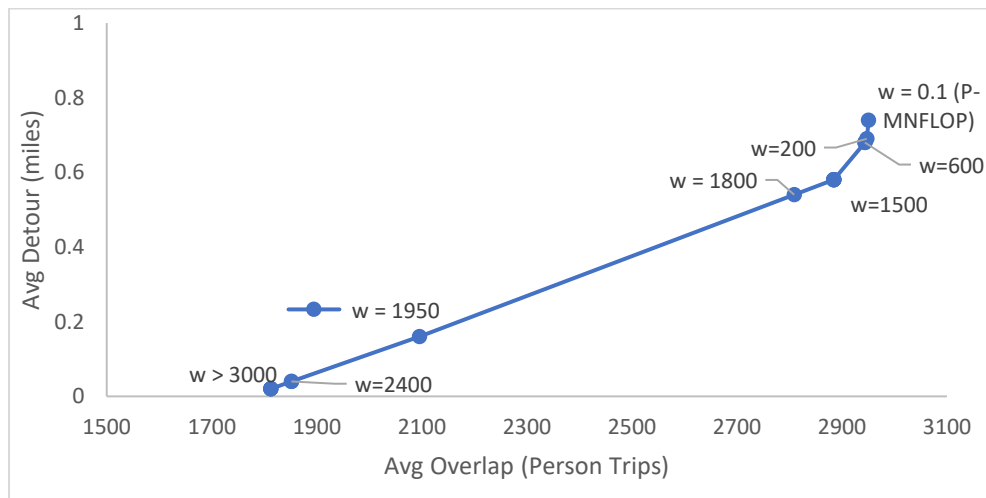

**Figure 7-6 Pareto Optimal Solutions for Scenario-1 Sioux Falls**

Appendix A includes further results for Sioux Falls Scenario Set 1, including Table A-1 that displays overlap/shareability metrics and path choice for every OD pair for all three assignment methods. Additionally, Figure A.0-1 in the Appendix shows the change in overlap percentage at the network level for different values of *w*.

## 7.2    Scenario Set 2: Origin-level Shareability Analysis

Scenario Set 2 analyzes the extent of overlaps between trips originating from each node of the Sioux Falls network by solving P-MNFLOP as well as the SP assignment problem, for each origin separately. Moreover, for each scenario (i.e., each origin), we only consider the trips originating from said origin. This analysis gives insights into the relative dispersion of trips originating from an individual location. Overlap percentage for an origin node indicates the percentage of total demand originating from the node with which a unit person-trip originating from the same node shares its path.

Figure 7-7 and Figure 7-8 display the magnitude of demand along with the overlap percentage for all 24 origin demand nodes in the Sioux Falls network using P-MNFLOP and SP, respectively. The plots

show that assigning trips using P-MNFLOP increases the overlap percentage for trips originating from each origin node when compared to SP assignment. Overlap percentage for origin nodes such as Node-1 and Node-20 nearly double when trips are assigned based on P-MNFLOP compared to SP. On the other hand, there is only a small increase in overlap percentage for trips originating from nodes such as Node-3, Node-9, Node-10, Node-17 and Node-19.

Figure 7-7 and Figure 7-8 also show that trips originating from Node-2 in the Sioux Falls network exhibit the highest overlap percentage (i.e., the lowest dispersion) under both P-MNFLOP and SP assignment. Node-2 has a flow overlap percentage of over 60% for P-MNFLOP. This means that when paths are assigned based on results of P-MNFLOP, the average trip starting from Node-2 shares its paths with 60% of all other person-trips originating from Node-2. On the other hand, trips originating from Node-10 have the lowest overlap percentage—21% for P-MNFLOP—despite having a high magnitude of originating trips.

Figure 7-7 and Figure 7-8 further reinforces that both the magnitude (indicated by total demand per unit time or total demand per unit area per unit time) as well as the direction of demand (indicated by percentage of overlap) are necessary to characterize of the shareability of trips from an origin location. Node-2 has a low magnitude of originating demand (4,000 person-trips), but a high overlap percentage value of 62%, meaning that trips from this location are still highly shareable. This might suggest operating a shared mobility feeder mode from this location might be viable, despite the low magnitude of demand. On the other hand, Node-6 has a lower overlap percentage value of 26%, however, since the magnitude of demand from the same node is high (26,000 person-trips), the average trip from Node-16 still shares paths with nearly 7,000 person-trips (26% of 26,000 person-trips). This means that even though trips from Node-16 are more dispersed, a unit flow from Node-16 is still likely to share a path in part or whole with nearly three times as many other person-trips compared to a unit person-trip starting from Node-2.

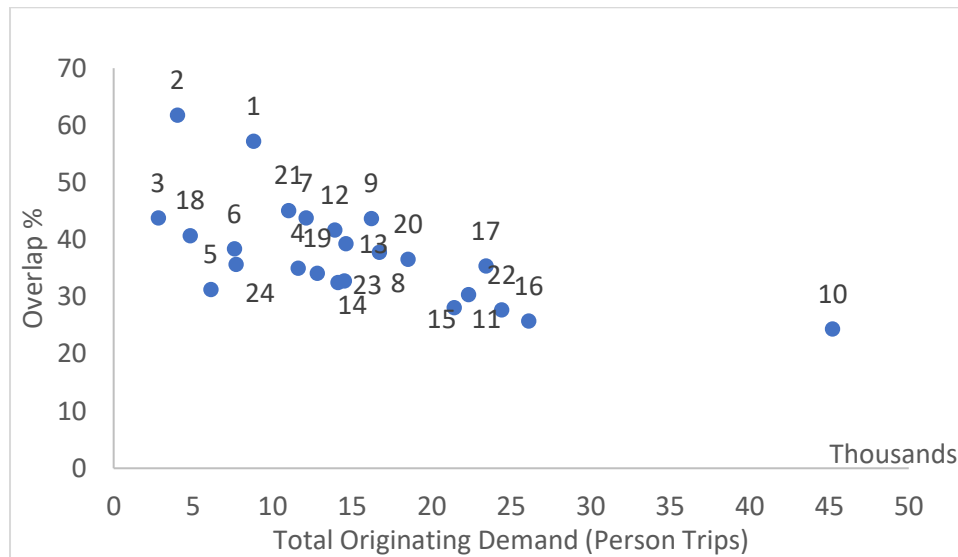

**Figure 7-7 Magnitude of Demand and Overlap Percentage for Origin Nodes in Sioux Falls with P-MNFLOP**

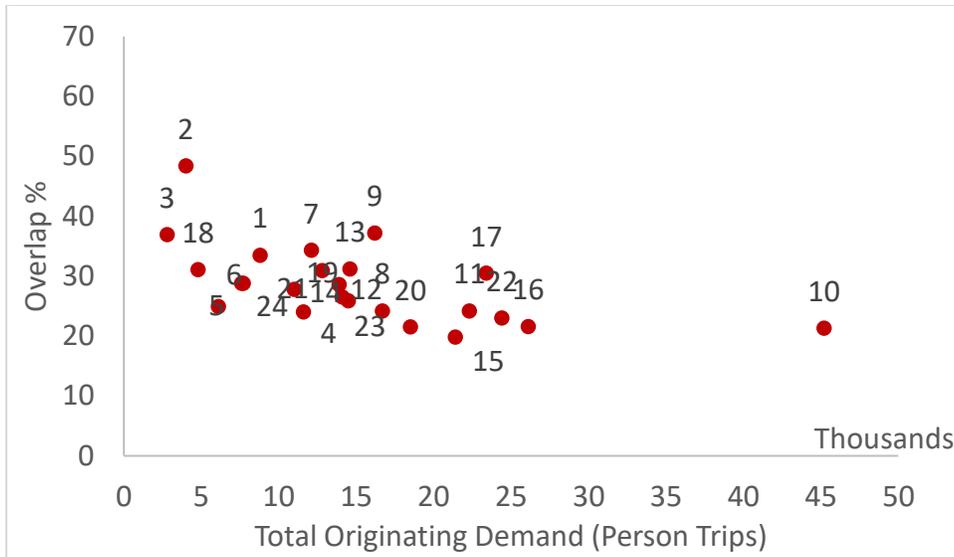

**Figure 7-8 Magnitude of Demand and Overlap Percentage for Origin Nodes in Sioux Falls with SP**

Figure 7-9 and Figure 7-10 display the link flows using P-MNFLOP for trips originating from Node-2 and Node-5, respectively. These two origin nodes have a similar magnitudes of total originating demand, whereas the overlap percentage for Node-2 is twice as much as Node-5. The higher value of overlap percentage for Node-2 is evident in Figure 7-9, where flows have a tree structure with a root at Node-2 and flows to destination nodes only use a few links/branches, indicating a high overlap between the paths to multiple destinations from the origin node. On the other hand, Figure 7-10 shows a tree structure with flows emanating from Node-5 branching out across many links on the way to destination nodes.

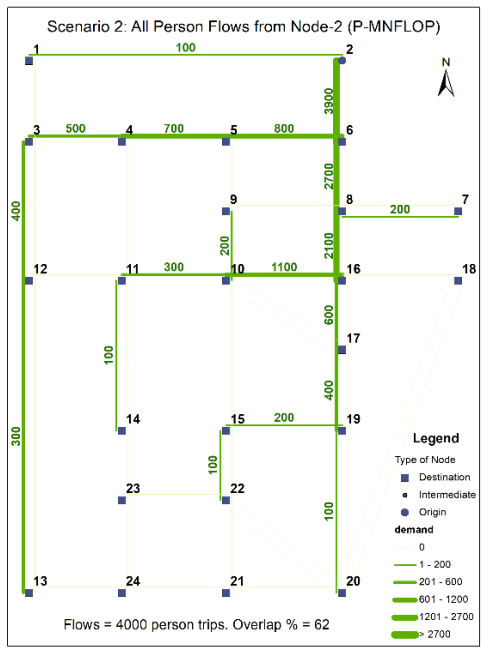

**Figure 7-9 P-MNFLOP Assignment from Node 2**

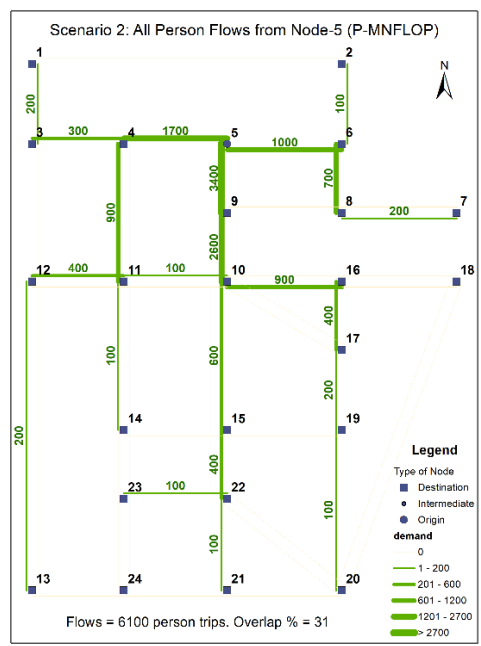

**Figure 7-10 P-MNFLOP Assignment from Node 5**

## 7.3    Scenario Set 3: Same Origin, Different Destinations Shareability Analysis

This scenario set aims to highlight the difference in shareability for different trip patterns for the same (or similar) magnitude of total demand originating from the same set of origin locations in the Sioux Falls network. The Northern nodes in the Sioux Falls network (nodes 1 to 6) are the origin nodes for all cases in this scenario set. A total demand of 1000 person-trips per unit time from these origin nodes is assigned separately to three different sets of destination nodes to show how shareability depends on the spatial distribution of demand.

The demand between each OD pair in each case is obtained by first slicing the original OD table to include only the selected OD pairs relevant to each case. The remaining OD table cells are all multiplied by a single factor such that the total demand of the OD table adds up to 1000 trips. Then we round the OD flows to the nearest whole number. Hence, the total demand may vary slightly from 1000 trips across the scenarios. This small variation does not affect the shareability metrics or inferences drawn.

Table 7.2 provides the shareability metric values for the three cases in this scenario set obtained using P-MNFLOP. The first case (N to CBD) has trip patterns from origin nodes in the north to the CBD (nodes 10 and 16), as shown in Figure 7-11. The second case shows trip patterns from origin nodes in the North to destination nodes in the South, as shown in Figure 7-12. In the final case, trips from origin nodes in the North are assigned randomly to destination nodes scattered throughout the network, as show in Figure 7-13.

**Table 7.2 Network Level Shareability Metrics for Sioux Falls (P-MNFLOP), Scenario Set 3**

| Network level Shareability Metric | N to CBD | N to S | N to Scatter |
|---|---|---|---|
| Total Demand | 1001 | 1008 | 999 |
| # Origins | 6 | 6 | 6 |
| # Destinations | 2 | 6 | 6 |
| Avg Overlap (Person-trips) | 566.3 | 461.1 | 305.2 |
| Avg Trip Distance (Miles) | 12.98 | 16.36 | 13.96 |
| Avg Overlap Distance (Miles) | 7.3 | 7.3 | 3.9 |
| Avg Overlap (%) | 56.63 | 46.02 | 30.58 |
| Average Detour (Miles) | 0.58 | 0.73 | 1.67 |
| Average Detour (Ratio) | 1.05 | 1.05 | 1.14 |
| Marginal Overlap (Trips/Detour mile) | 272.6 | 262.7 | 51 |
| Marginal Overlap % (Overlap %/Detour mile) | 27.3 | 26.2 | 5.1 |
| Detour Distance Elasticity of Overlap | 8.3 | 15.2 | 2.9 |
| Average Link Flow (Person-trips) | 342 | 228 | 164 |
| # Links with non-zero Flows | 10 | 21 | 22 |

The results in Table 7.2 indicate that the magnitude of overlap, as well as overlap percentage is the highest when trips from origin nodes are sent to a few destination nodes that are close to each other (i.e., the N to CBD case). The number of overlapping trips and overlap percentage are the lowest when trips are bound to destination nodes scattered throughout the network (N to Scatter case). The higher marginal values of overlap for a unit detour for the first two cases compared to the third case further indicates that when trip patterns are concentrated onto fewer destinations that are close to each other, even a small detour from the shortest path can result in significant gains in the potential to share paths with other trips. The ratio of average overlap distance to average total trip distance is highest for the N to CBD case followed by N to S and the N to Scatter cases. This indicates that a unit trip in the N to CBD case overlaps with the largest fraction of other flows in the network along each link in its path. This is again due to trip destinations being clustered, enabling more overlaps. The highest overlap and overlap percentage values for the N to CBD case are also reflected in its high value of average link flows and the number of links used, indicating that P-MNFLOP is able to maximize overlap by concentrating more flows onto fewer links in this case. Even though the number of links with non-zero flows is similar in both the N to S and N to Scatter cases, the average link flows on them are higher for the former, since the destinations are more clustered compared to the latter case. Even though N to CBD case has the highest value of overlap and overlap percentage, as well as their marginal values, it does not have the highest value for detour elasticity of overlap. This is because of the high base overlap value in the N to CBD case when flows are assigned using SP assignment instead of P-MNFLOP. With a detour distance elasticity of overlap value of 15.2, P-MNFLOP results in the most improvement in overlapping flows in the N to S case compared to when trips are assigned on the shortest paths.

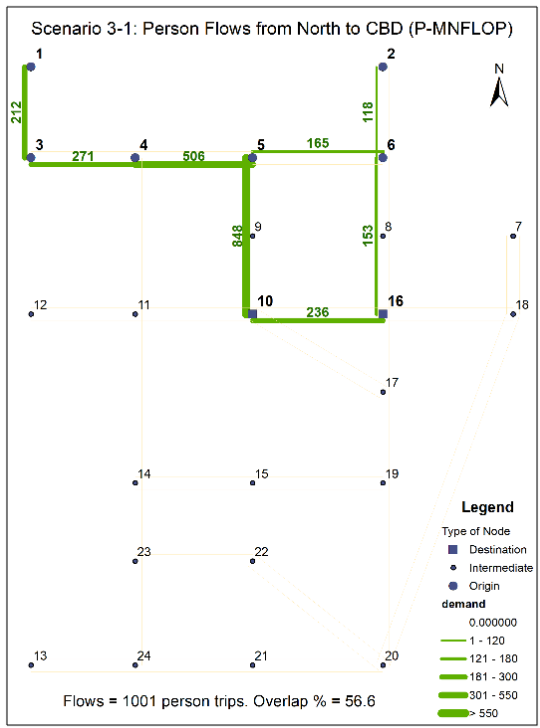

**Figure 7-11 Link Flows using P-MNFLOP, Sioux Falls Scenario 3-1**

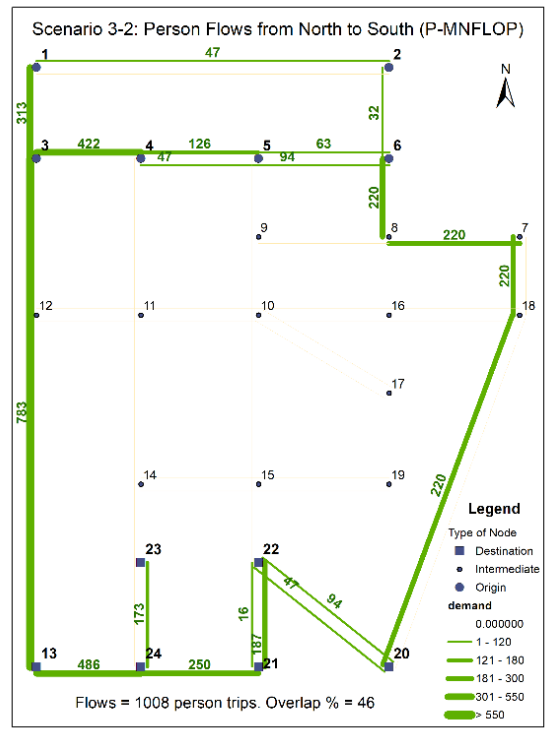

**Figure 7-12 Link Flows using P-MNFLOP, Sioux Falls Scenario 3-2**

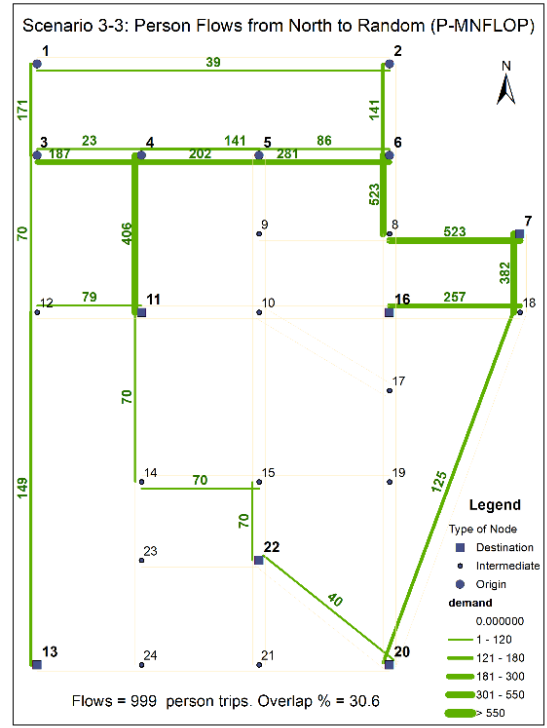

**Figure 7-13 Link Flows using P-MNFLOP, Sioux Falls Scenario 3-3**

## 7.4 Scenario Set 4: Different Origins, Different Destinations Shareability Analysis

This scenario set aims to highlight the difference in shareability metrics when the same (or similar) magnitude of total demand in the network has different underlying trip patterns. Unlike the prior scenario set, in Scenario Set 4 both the origins and destinations vary across the four cases. This section uses the same process as Scenario Set 3 for generating an OD table for each scenario. Once again, the OD table for each case has around 1000 trips.

The first case (case 4-1) involves trips assigned from non-CBD nodes toward CBD nodes (Nodes 10 and 16). The second and third cases (4-2 and 4-3) involve trips assigned between random OD pairs scattered throughout the Sioux Falls network. The fourth case (4-4) has trips assigned between a set of clustered origin nodes and a set of clustered destination nodes.

Table 7.3 outlines the shareability metrics for all 4 cases in this scenario set. The person-trips on links in each of the 4 cases are shown in Figure 7-14, Figure 7-15, Figure 7-16, and Figure 7-17 respectively. Results show that, similar to what was observed in the previous subsection, the magnitude and percentage of overlapping trips are comparatively higher in cases where trips either originate from and/or terminate at nodes that are either fewer or closer to each other. Case 4-4 (Figure 7-17) has the highest magnitude (493 trips) as well as percentage (49.5%) of overlapping trips because in this case both the origin nodes and the destination nodes are clustered (origin nodes in the South-West region and destination nodes in the North-East region of the network). Clustered origins and destinations allow P-MNFLOP assignment to increase overlapping flows by assigning trips to paths with a very small detour (average detour of 7% of shortest path trip distance). This is further corroborated by the high distance elasticity of overlap observed in this case (11%). The high ratio of average overlap distance to average trip detour distance in this case also reflects the fact that a unit trip shares paths with a high fraction of all flows in the network on each link in its path. This case also has the highest average link flows as well as the least number of used links (links with non-zero flows), indicating that P-MNFLOP is able to assign more flows onto fewer links and fewer corridors.

Case 4-1 (Figure 7-14) has the 2$^{nd}$ highest magnitude and percentage of overlaps because even though the origin nodes are scattered all over the region, all trips are bound to the CBD nodes (Node-10 and Node-16) that are close to each other. Case 4-2 (Figure 7-15) and 4-3 (Figure 7-16) have low shareability for the same magnitude of total demand compared to the other two cases since their origin and destination nodes are scattered throughout the network. It is also interesting to note that even though Case 4-1 has a higher value for overlapping trips as well as overlap % compared to Case 4-2 and Case 4-3, it has lower values for marginal overlap, marginal overlap % and distance elasticity of overlap. This is because P-MNFLOP is not able to significantly increase overlapping flows in Case 4-1 compared to when all flows are assigned on shortest paths in this case. A 1% increase in trip distance yielded only a 2.5% increase in overlapping flows for Case 4-1 – the lowest value among all cases in this scenario set. Case 4-1 also has a high number of links used (links with non-zero flows), which could be explained by the spatial extent of origin nodes which are scattered all over the network in this case.

**Table 7.3 Network Level Shareability Metrics (P-MFNLOP) for Sioux Falls Scenario Set-4**

| Network level Shareability Metric | 4-1 Outskirts to CBD | 4-2 Random Case 1 | 4-3 Random Case 2 | 4-4 Clustered O's to Clustered Ds |
|---|---|---|---|---|
| **Total Demand** | 998 | 1000 | 995 | 998 |
| **# Origins** | 22 | 5 | 10 | 5 |
| **# Destinations** | 2 | 8 | 5 | 5 |
| **Avg Overlap (Trips)** | 178.4 | 115.3 | 144.3 | 492.9 |
| **Avg Overlap (%)** | 17.89 | 11.54 | 14.51 | 49.5 |
| **Avg Trip Distance (Miles)** | 9.33 | 7.11 | 9.15 | 15.09 |
| **Avg Overlap Dist. (Miles)** | 1.6 | 0.8 | 1.3 | 7.1 |
| **Average Detour (Miles)** | 0.82 | 0.4 | 1.01 | 0.97 |
| **Average Detour (Ratio)** | 1.1 | 1.06 | 1.12 | 1.07 |
| **Marginal Overlap (Trips/Detour mile)** | 42.3 | 58 | 42.7 | 219.6 |
| **Marginal Overlap % (Overlap %/Detour mile)** | 4.2 | 5.8 | 4.3 | 22 |
| **Distance Elasticity of Overlap** | 2.5 | 4.2 | 3.4 | 11.1 |
| **Average Link Flow (Person-trips)** | 71 | 72 | 67 | 307 |
| **# Links with non-zero flows** | 34 | 25 | 37 | 14 |

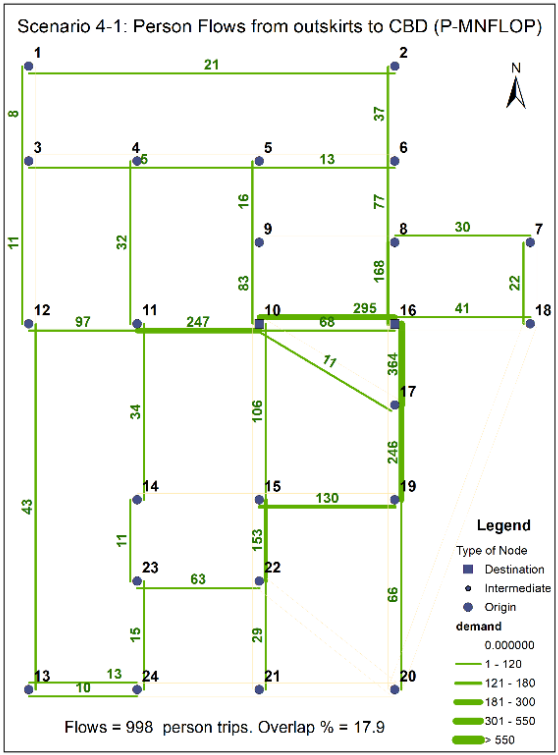

**Figure 7-14 Link Flows for Sioux Falls Scenario 4-1**

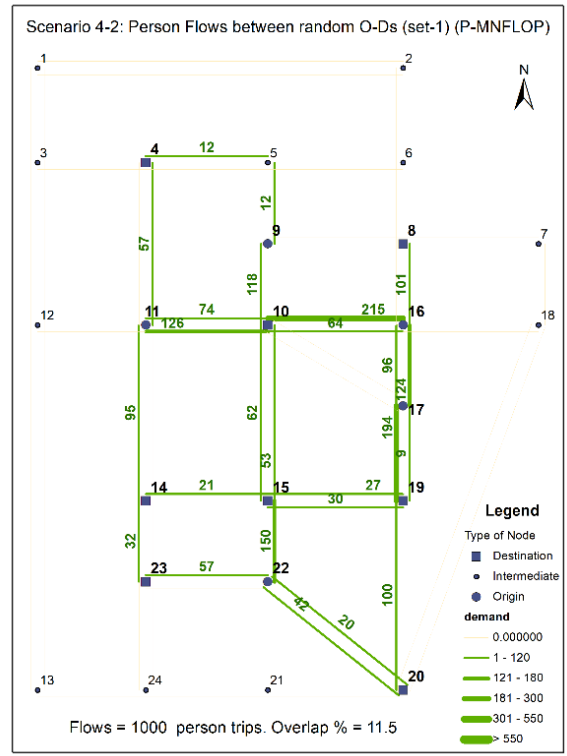

**Figure 7-15 Link Flows for Sioux Falls Scenario 4-2**

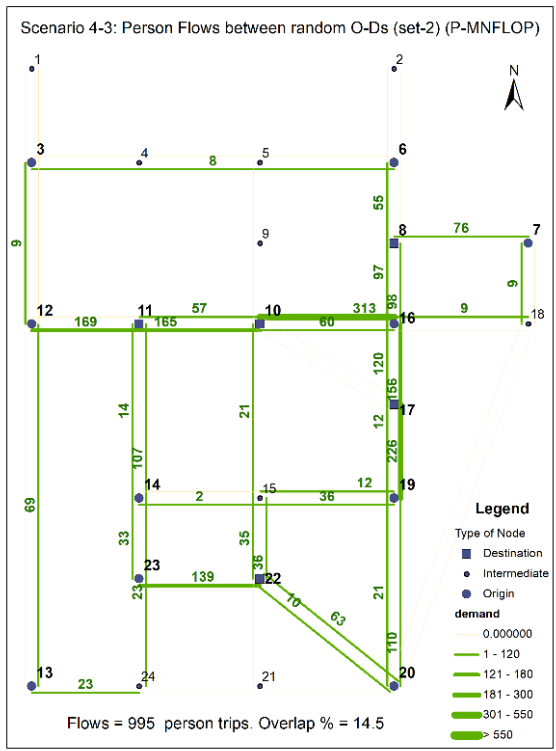

**Figure 7-16 Link Flows for Sioux Falls Scenario 4-3**

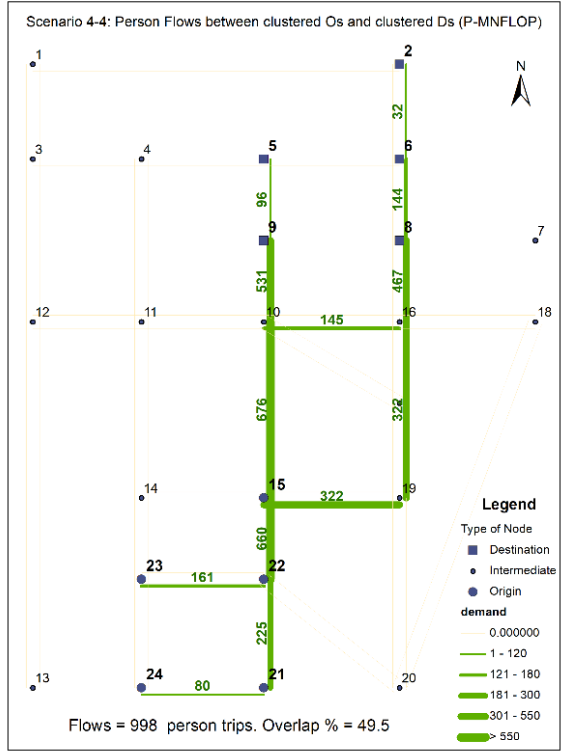

**Figure 7-17 Link Flows for Sioux Falls Scenario 4-4**

## 7.5   Discussion

The results in the prior 4 subsections show that (i) P-MNFLOP and Bi-MNFLOP models and a commercial solver can significantly increase overlaps in a transportation network compared to shortest path assignment; (ii) MNFLOP and its associated shareability metrics capture the relative dispersion/concentration of trips emanating from/to and origin/destination node; and (iii) MNFLOP and its associated shareability metrics can effectively differentiate between the shareability potential of the same network under different OD trip tables, wherein clustered origins and/or clustered destinations increase flow overlaps and shareability potential relative to the same total demand scattered across the network.

The ability of MNFLOP and its associated metrics to meet each of these three objectives, indicates that it can be used to differentiate between cities, subregions with cities, and even locations within a city in terms of their shareability potential. The next section, in addition to summarizing the study and discussing limitations, provides an extensive discussion of the potential uses of MNFLOP and its shareability metrics proposed in this paper.

# 8   Conclusion

## 8.1   Summary

This paper contains a novel approach to quantify shareability of person-trips in a region, given OD travel demand and an underlying transportation network. The study conceptualizes and defines mathematically the notion of 'flow overlaps' for a unit person-trip in a transportation network. The paper employs the flow overlap concept to formulate the MNFLOP, a path- and flow- based Quadratic Integer Program, to find the shareability for a given network and OD demand. A solution to the MNFLOP includes the optimal paths between each OD pair that maximize shareability in the network. The paper then uses the MNFLOP output to calculate many shareability metrics at various levels of aggregation: OD, origin, link, and network level. The study uses the MNFLOP model to analyze shareability in the Sioux Falls network, under multiple demand scenarios. The origin-level metrics of shareability proposed in this paper add another dimension to measuring demand apart from its magnitude, specifically the dispersion (or overlap percentage) of trips from a node/location capture the directional component of demand.

The computational results verify the MNFLOP and associated shareability metrics can meaningfully distinguish between the shareability of a given street network across different spatial distributions of demand. The results also indicate that slight increases in detour for some OD pairs, compared to their shortest paths, can significantly increase flow overlaps in a network.

## 8.2   Limitations

One shortcoming of the existing study is the reliance on a commercial solver for the quadratic integer programming model. Future research should work to develop alternative exact and heuristic solution algorithms to solve the problem for larger networks, as well as consideration of alternative formulations and approximations of the original model. Approximations include reducing the problem size by clustering trip origin and destination locations. Another approximation includes restricting the other candidate OD pairs that constitute overlapping flows when calculating overlap for a specific OD pair, to reduce the number of quadratic terms. In terms of heuristic solution approaches, genetic algorithms are potentially appealing as they can explore the solution space effectively and efficiently. Tailored heuristics that efficiently estimate, rather than fully evaluate, the objective function in sub-iterations may also be effective.

Another shortcoming is that the results of the MNFLOP model depend on the possible paths for each OD pair and the method used to generate the set of $k$ paths. Future research can explore running the MNFLOP model using a range of methods to generate $k$ paths for each OD pair, to see how dependent the results are on path generation.

Another shortcoming is the lack of consideration of link capacities in the model. Despite wanting to maximize flow overlaps in the network by condensing flows onto fewer links in order to quantify sharing potential, in the real-world links in a transportation network do have maximum capacities. To incorporate link capacity, one might consider the maximum person throughput of a heavy rail line, if it is feasible for the network link to be converted into a heavy rail line. Another option is to determine the maximum person throughput of a bus rapid transit line and use this value to constrain flow on each link in the network.

## 8.3   Future Research

This subsection includes a rather extensive discussion of future research directions related to MNFLOP and associated shareability metrics. The reason being that we believe MNFLOP and the shareability metrics presented in this paper have significant potential to provide insights into many transportations system planning and design decisions. While this paper clearly *verifies* that the MNFLOP model and shareability metrics can (i) assign OD person-trips to network paths that maximize flow overlap, (ii) measure and differentiate the shareability potential of different OD demand patterns in a given network, and (iii) measure and differentiate demand dispersion from single nodes/locations in a network, the paper does not yet *validate* the usefulness of MNFLOP and its associated shareability metrics. Nevertheless, the MNFLOP and shareability metrics should have considerable value in transportation system planning and design, as described in the next two subsections.

### 8.3.1   Shareability as Input for Shared Mobility System Design

The original impetus for this study was that the existing flexible transit design literature did not effectively characterize demand dispersion alongside demand density in a service, nor did the existing literature consider the region's underlying transportation network. The authors first sought to determine the optimal shared mobility mode (i.e., fixed-route transit, flexible transit, ridesharing, or ridesourcing) or mode combination for each area of a city given the properties of the area (e.g., demand density and dispersion and underlying transportation network). Although this study does not attempt to link demand density, demand dispersion, and network structure to the viability of specific shared mobility modes through MNFLOP and shareability metrics, ongoing research does attempt to make this linkage. If this linkage exists, it would provide considerably more value to transportation system planners than existing linkages between homogeneous demand density in a service area, and the viability of shared mobility modes.

Considering the study's background, the authors believe the MNFLOP and associated shareability metrics can provide valuable insights for planning multi-modal shared mobility networks and even function as an input to planning and design models for multi-modal shared mobility networks. While researchers and practitioners understand the broad generalizations that (i) high-capacity, high-frequency fixed-route transit services work well in dense areas, and (ii) flexible, demand-adaptive door-to-door services work better in lower density areas, real-world urban areas and networks do not fit neatly into these two extremes. Hence, deciding (i) where to operate specific shared mobility modes ranging from heavy rail to ridesourcing and (ii) the optimal combination of shared mobility modes in a city, are ongoing challenges that the MNFLOP and associated shareability metrics can help address.

In addition to helping assess the viability of specific shared mobility modes, we believe the MNFLOP output, specifically the network visualizations with link flows (e.g., Figure 7-1-Figure 7-3, and Figure 7-14-Figure 7-17), can identify the street segments and corridors wherein high-capacity, high-frequency transit lines are most viable. The link level shareability metrics should have particular value in identifying streets and corridors to locate high-frequency, high-capacity transit infrastructure and services. Moreover, the authors believe the bi-criteria MNFLOP can have particular value for multi-modal system design as it captures trade-offs between overlapping paths and required person-trip detours.

Finally, the origin level analysis and overlap percentage metric proposed in this study capture the magnitude and the extent of directional overlap of demand starting from an origin. Having a high percentage of flow overlap from a node, notwithstanding a low magnitude, indicates that it is possible to operate shared modes even in low density areas when there is high spatial and temporal overlap of flows starting from the location. These origin-level shareability metrics should provide considerable value when planning first- and last-mile feeder services around transit stations.

### 8.3.2 Shareability Metrics as Predictors of Transit Ridership

Transit ridership and transit mode share vary across cities in the United States and the world. Moreover, demand for transit varies across spatial areas within cities. Similarly, the demand for ridesourcing and shared-ride mobility-on-demand services vary across and within cities. Transit operators, mobility service providers, and transportation planners are quite interested in understanding the factors that cause these variations in transit usage across and within cities. The authors believe that the shareability metrics presented in this study can be used, along with various other factors, to explain and forecast the usage of shared mobility modes like public transit and shared-ride services. After controlling for sociodemographic attributes of urban areas as well as transit expenditures and other factors, cities with higher shareability are likely to have higher transit ridership. Future research can test this hypothesis.

## Acknowledgement

The first author received partial funding for this research from the Gordon Hein Scholarship in 2020-2021. The authors would like to express their gratitude to the scholarship sponsor as well as the Graduate Division at University of California Irvine for awarding the scholarship.

The authors have no competing interests to declare.

# Appendix A: Additional Results for Sioux Falls Scenario – 1

Table A-1 displays each OD pair, the demand for each OD pair, as well as the optimal path, trip overlap, overlap percentage, detour distance, and marginal overlap for three separate assignment methods—P-MNFLOP, SP, and Bi-MNFLOP.

**Table A- 1 Comparison of OD Level Overlap Metrics for Sioux Falls Scenario-1**

| OD Pair | Demand ($F_{od}$) | k | Optimal Path | P-MNFLOP Trip Overlap ($Z_{od}$) | Overlap % ($Z^{\%}_{od}$) | Δ (Miles) | Marginal Overlap ($M_{Z_{od}}$) (Trips/Mile) | SP Trip Overlap ($Z_{od}$) | Overlap % ($Z^{\%}_{od}$) | Bi-MNFLOP (w=1800) Trip Overlap ($Z_{od}$) | Overlap % ($Z^{\%}_{od}$) |
|---|---|---|---|---|---|---|---|---|---|---|---|
| **1-13** | 500 | 0 | 1-3-12-13 | 3908.1 | 61.07 | 0 | - | 2780.8 | 43.46 | 3844.5 | 60.08 |
| **1-20** | 300 | 4 | 1-3-12-13-24-21-22-20 | 2563 | 40.05 | 3 | 522.8 | 994.5 | 15.54 | 2536.5 | 39.64 |
| **1-21** | 100 | 0 | 1-3-12-13-24-21 | 3343.4 | 52.25 | 0 | - | 2165.7 | 33.84 | 3282.3 | 51.29 |
| **1-22** | 400 | 0 | 1-3-12-13-24-21-22 | 3129 | 48.9 | 0 | - | 1999 | 31.24 | 3044 | 47.57 |
| **1-23** | 300 | 0 | 1-3-12-13-24-23 | 3387.2 | 52.93 | 0 | - | 2199 | 34.36 | 3310.8 | 51.74 |
| **1-24** | 100 | 0 | 1-3-12-13-24 | 3692.3 | 57.7 | 0 | - | 2439 | 38.12 | 3619 | 56.56 |
| **2-13** | 300 | 0 | 2-1-3-12-13 | 2634.3 | 41.17 | 0 | - | 1904.9 | 29.77 | 2593.1 | 40.52 |
| **2-20** | 100 | 0 | 2-6-8-7-18-20 | 1024 | 16 | 0 | - | 1255.2 | 19.62 | 955.2 | 14.93 |
| **2-22** | 100 | 0 | 2-6-8-7-18-20-22 | 922.8 | 14.42 | 0 | - | 1027.6 | 16.06 | 822.8 | 12.86 |
| **3-13** | 100 | 0 | 3-12-13 | 4999 | 78.12 | 0 | - | 3399 | 53.12 | 4899 | 76.56 |
| **3-22** | 100 | 0 | 3-12-13-24-21-22 | 3411.5 | 53.31 | 0 | - | 2074 | 32.41 | 3305.2 | 51.65 |
| **3-23** | 100 | 0 | 3-12-13-24-23 | 3814.4 | 59.61 | 0 | - | 2352.8 | 36.77 | 3714.4 | 58.05 |
| **4-13** | 600 | 0 | 4-3-12-13 | 4162.6 | 65.05 | 0 | - | 2671.7 | 41.75 | 4062.6 | 63.49 |
| **4-20** | 300 | 0 | 4-5-6-8-7-18-20 | 1081.4 | 16.9 | 0 | - | 1163.7 | 18.19 | 969.6 | 15.15 |
| **4-21** | 200 | 0 | 4-3-12-13-24-21 | 3499 | 54.68 | 0 | - | 2099 | 32.8 | 3415.7 | 53.38 |
| **4-22** | 400 | 4 | 4-3-12-13-24-21-22 | 3269 | 51.09 | 2 | 1215.6 | 837.9 | 13.09 | 3164 | 49.45 |
| **4-23** | 500 | 3 | 4-3-12-13-24-23 | 3551.9 | 55.51 | 3 | 855.7 | 984.7 | 15.39 | 3451.9 | 53.95 |
| **4-24** | 200 | 0 | 4-3-12-13-24 | 3879 | 60.62 | 0 | - | 2359 | 36.87 | 3779 | 59.06 |
| **5-13** | 200 | 0 | 5-4-3-12-13 | 3645.2 | 56.96 | 0 | - | 2352.8 | 36.77 | 3545.2 | 55.4 |
| **5-20** | 100 | 0 | 5-6-8-7-18-20 | 1185.7 | 18.53 | 0 | - | 1279 | 19.99 | 1059 | 16.55 |
| **5-21** | 100 | 1 | 5-4-3-12-13-24-21 | 3229 | 50.46 | 1 | 2887.9 | 341.1 | 5.33 | 3144 | 49.13 |
| **5-22** | 200 | 2 | 5-6-8-7-18-20-22 | 1039 | 16.24 | 3 | 223.1 | 369.6 | 5.78 | 199 | 3.11 |
| **5-23** | 100 | 1 | 5-4-3-12-13-24-23 | 3262.2 | 50.98 | 3 | 775.2 | 936.5 | 14.64 | 3162.2 | 49.42 |
| **6-13** | 200 | 0 | 6-5-4-3-12-13 | 2881.4 | 45.03 | 0 | - | 1869.6 | 29.22 | 2781.4 | 43.47 |
| **6-20** | 300 | 0 | 6-8-7-18-20 | 1399 | 21.86 | 0 | - | 1599 | 24.99 | 1299 | 20.3 |
| **6-21** | 100 | 1 | 6-8-7-18-20-22-21 | 1032.3 | 16.13 | 1 | -72.5 | 1104.9 | 17.27 | 875.5 | 13.68 |
| **6-22** | 200 | 0 | 6-8-7-18-20-22 | 1149 | 17.96 | 0 | - | 1192.8 | 18.64 | 1017.8 | 15.9 |
| **6-23** | 100 | 4 | 6-5-4-3-12-13-24-23 | 2764.2 | 43.2 | 3 | 651.7 | 809 | 12.64 | 834 | 13.03 |
| **6-24** | 100 | 1 | 6-5-4-3-12-13-24 | 2922.8 | 45.68 | 1 | 1968.8 | 954 | 14.91 | 2822.8 | 44.11 |

Figure A.0-1 shows the change in overlap percentage at the network level for different values of $w$. The plot indicates that for lower values of $w$, the average overlap percentage in the network is the high, with the results of P-MNFLOP yielding the highest value of overlap percentage ($w = 0.1$). As the value of $w$ increases, the Bi-MNFLOP objective moves closer to a SP assignment resulting in lower overlap percentage with reducing detours. The plot indicates that the average overlap percentage in a network is a monotonically decreasing function of detour parameter $w$. Similar to the trend observed in Figure 7-6, the plot is not a smooth function, it includes sharp drops in overlap percentage over a small variation in $w$. This plot also indicates that a significant gain in shareability in a network can be achieved if person-trips in the network are incentivized to take even a slight detour from their SP.

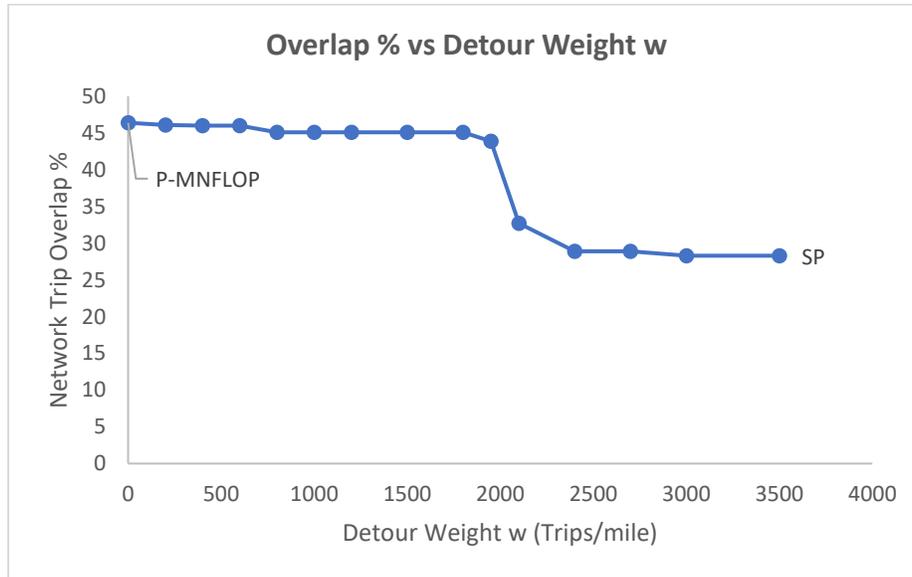

**Figure A.0-1 Overlap Percentage vs Detour Weight w for Bi-MNFLOP assignment**